\pdfoutput=1
\documentclass[12pt]{article}
\usepackage{graphicx}
\usepackage{subcaption}
\usepackage{amssymb}
\usepackage{amsmath}
\usepackage{amsfonts}
\usepackage{multirow}

\usepackage{xcolor}
\usepackage{url}
\usepackage{ulem}
\usepackage{float}
\usepackage{float}

\usepackage{pifont}

\usepackage{soul}
\usepackage[english]{babel}
\usepackage[T1]{fontenc}
\usepackage{lmodern}
\usepackage[utf8]{inputenc}  
\usepackage{bbm}

\usepackage{hyperref}
\hypersetup{
  colorlinks   = true, 
  urlcolor     = blue, 
  linkcolor    = black, 
  citecolor   = black 
}

\usepackage[numbers,sort&compress]{natbib}
\graphicspath{{plots/}}

\newcommand{\lsim}
{\;\raisebox{-.3em}{$\stackrel{\displaystyle <}{\sim}$}\;}
\newcommand{\gsim}
{\;\raisebox{-.3em}{$\stackrel{\displaystyle >}{\sim}$}\;}

\newcommand\tb{\tan\beta}

\newcommand\LP{\left(}
\newcommand\RP{\right)}

\newcommand\ra{\rightarrow}
\newcommand\tenp[1]{\times 10^{#1}}

\newcommand\ReDiag{\mathop{%
  \raise .5pt\hbox{[}%
  \widetilde{\mathrm{Re}}%
  \raise .5pt\hbox{]}}}
\newcommand\ReOffDiag{\mathop{%
  \raise .5pt\hbox{$\llbracket$}%
  \widetilde{\mathrm{Re}}%
  \raise .5pt\hbox{$\rrbracket$}}}

\newcommand\DRbar{\ensuremath{\smash{\overline{\mathrm{DR}}}}}

\newcommand\MW{M_W}
\newcommand\MZ{M_Z}
\newcommand\Mh{M_h}

\newcommand\MA{M_A}

\newcommand\Sn{\tilde\nu}
\newcommand\Sl{\tilde l}
\newcommand\sle[1]{\tilde l_{#1}}
\newcommand\Slpm{\tilde l^\pm}

\newcommand\Sel[1]{\tilde e_{#1}}
\newcommand\Smu[1]{\tilde \mu_{#1}}

\newcommand\mse[1]{m_{\Sel{#1}}}
\newcommand\msl[1]{m_{\Sl_{#1}}}

\newcommand\Stau[1]{{\tilde\tau_{#1}}}
\newcommand\stau{\tilde \tau}

\newcommand\mL{m_{\tilde l_L}}
\newcommand\mR{m_{\tilde l_R}}

\newcommand\ino[1]{\tilde\chi_{#1}}

\newcommand\chapm[1]{\ino{#1}^\pm}

\newcommand\cha{\chapm}
\newcommand\mcha[1]{m_{\chapm{#1}}}

\newcommand\neu[1]{\ino{#1}^0}
\newcommand\mneu[1]{m_{\neu{#1}}}

\newcommand\refeq[1]{Eq.~(\ref{#1})}
\newcommand\refeqs[1]{Eqs.~(\ref{#1})}
\newcommand\refta[1]{Tab.~\ref{#1}}
\newcommand\refse[1]{Sect.~\ref{#1}}

\newcommand\citere[1]{Ref.~\cite{#1}}
\newcommand\citeres[1]{Refs.~\cite{#1}}

\newcommand{\CP}{{\cal CP}}
\newcommand{\cp}{{\CP}}

\newcommand{\tev}{\,\, \mathrm{TeV}}
\newcommand{\gev}{\,\, \mathrm{GeV}}
\newcommand{\mev}{\,\, \mathrm{MeV}}

\newcommand\MO{\texttt{MicrOMEGAs}}
\newcommand\CM{\texttt{CheckMATE}}

\newcommand\fb{\ensuremath{\mbox{fb}}}
\newcommand\ab{\ensuremath{\mbox{ab}}}

\newcommand\ifb{\ensuremath{\fb^{-1}}}
\newcommand\iab{\ensuremath{\ab^{-1}}}

\newcommand\msmu[1]{m_{\tilde{\mu}_{#1}}}
\newcommand\mstau[1]{m_{\tilde{\tau}_{#1}}}

\newcommand{\br}{\text{BR}}

\newcommand{\sig}{\sigma}

\def\order#1{\ensuremath{{\cal O}(#1)}}
\def\reffi#1{\mbox{Fig.~\ref{#1}}}
\def\reffis#1{\mbox{Figs.~\ref{#1}}}

\def\ga{\gamma}
\def\De{\Delta}

\def\gmin2{\ensuremath{(g-2)_\mu}}
\def\amu{\ensuremath{a_\mu}}
\def \met  {\mbox{${E\!\!\!\!/_T}$}}
\newcommand{\ssi}{\ensuremath{\sig_p^{\rm SI}}}

\definecolor{Orange}{named}{orange}
\definecolor{Purple}{named}{purple}
\definecolor{Lightblue}{cmyk}{0.9,0.1,0.1,0.3}
\definecolor{dgelborange}{cmyk}{0.,0.3,0.5, 0.}
\definecolor{Lila}{rgb}{0.5,0.,1}
\definecolor{Darkgreen}{rgb}{0.,.7,0.2}

\evensidemargin \oddsidemargin
\allowdisplaybreaks
\begin{document}
\thispagestyle{empty}

\def\thefootnote{\fnsymbol{footnote}}

\begin{flushright}
\mbox{}
IFT--UAM/CSIC--21-022\\
IPMU21-020 
\end{flushright}

\vspace{0.3cm}

\begin{center}

{\large\sc 
{\bf Improved \boldmath{\gmin2} Measurements\\[.5em]
and Wino/Higgsino Dark Matter
}}

\vspace{0.7cm}

{\sc
Manimala Chakraborti$^{1}$%
\footnote{email: mani.chakraborti@gmail.com}%
, Sven Heinemeyer$^{2,3,4}$%
\footnote{email: Sven.Heinemeyer@cern.ch}%
~and Ipsita Saha$^{5}$%
\footnote{email: ipsita.saha@ipmu.jp}%
}

\vspace*{.5cm}

{\sl
$^1${Astrocent, Nicolaus Copernicus Astronomical Center of the Polish Academy of Sciences,
ul. Rektorska 4, 00-614 Warsaw, Poland\\}

\vspace*{0.1cm}

$^2$Instituto de F\'isica Te\'orica (UAM/CSIC), 
Universidad Aut\'onoma de Madrid, \\ 
Cantoblanco, 28049, Madrid, Spain

\vspace*{0.1cm}

$^3$Campus of International Excellence UAM+CSIC, 
Cantoblanco, 28049, Madrid, Spain 

\vspace*{0.1cm}

$^4$Instituto de F\'isica de Cantabria (CSIC-UC), 
39005, Santander, Spain
\vspace*{0.1cm}

$^5$Kavli IPMU (WPI), UTIAS, University of Tokyo, Kashiwa, Chiba 277-8583, Japan
}

\end{center}

\vspace*{0.1cm}

\begin{abstract}
\noindent
The electroweak (EW) sector of the Minimal Supersymmetric Standard Model
(MSSM) can account for variety of experimental data.
In particular it can explain the persistent $3-4\,\sig$ discrepancy
between the experimental result for the anomalous magnetic moment of the
muon, \gmin2, and its Standard Model (SM) prediction. 
The lighest supersymmetric particle (LSP), which we take as the lightest
neutralino, $\neu1$, can furthermore account for the observed Dark Matter (DM) 
content of the universe via coannihilation with the next-to-LSP
(NLSP), while being in agreement with negative results from
Direct Detection (DD) experiments. Concerning the unsuccessful searches
for EW particles at the LHC, owing to relatively small production
cross-sections a comparably light EW sector of the MSSM is in full 
agreement with the experimental data.
The DM relic density can fully be explained by a mixed bino/wino LSP.
Here we take the relic density as an upper bound, which opens up the
possibility of wino and higgsino DM. We first analyze which mass ranges of
neutralinos, charginos and scalar leptons are in agreement with all
experimental data, including relevant LHC searches.
We find roughly an upper limit of $\sim 600 \gev$ for
the LSP and NLSP masses. 
In a second step we assume that the new result of the 
Run~1 of the ``MUON G-2'' collaboration at Fermilab yields a precision
comparable to the existing experimental result with the same central
value. We analyze the potential impact of the combination of the Run~1
data with the existing \gmin2\ data on the allowed
MSSM parameter space. We find that in this case the upper limits
on the LSP and NLSP masses are substantially reduced by roughly
$100 \gev$. We interpret these upper bounds in view of future HL-LHC
EW searches as well as future high-energy
$e^+e^-$~colliders, such as the ILC or CLIC. 
\end{abstract}


\def\thefootnote{\arabic{footnote}}
\setcounter{page}{0}
\setcounter{footnote}{0}

\newpage


\section{Introduction}
\label{sec:intro}

One of the most important tasks at the LHC is to search for physics beyond the 
Standard Model (SM). This includes the production 
and measurement of the properties of Cold Dark Matter (CDM). 
These two (related) tasks will be among the top priority in the future
program of high-energy particle physics. One tantalizing hint for
physics beyond the SM (BSM) is the anomalous magnetic moment of the
muon, \gmin2. The experimental result deviates
from the SM prediction by
$3-4\sig$~\cite{Keshavarzi:2019abf,Davier:2019can}. 
Improved experimental results are expected soon~\cite{gmt-new} from the 
Run~1 data of the ``MUON G-2'' experiment~\cite{Grange:2015fou}.  
Another clear sign for BSM physics is the precise measurement of the CDM relic
abundance~\cite{Planck}. A final set of related constraints comes from
CDM Direct Detection (DD) experiments. The LUX~\cite{LUX},
PandaX-II~\cite{PANDAX} and XENON1T~\cite{XENON} experiments provide
stringent limits on the spin-independent (SI) DM scattering cross-section, \ssi.

\smallskip
Among the BSM theories under consideration the Minimal Supersymmetric
Standard Model  
(MSSM)~\cite{Ni1984,Ba1988,HaK85,GuH86} is one of the leading candidates.
Supersymmetry (SUSY) predicts two scalar partners for all SM fermions as well
as fermionic partners to all SM bosons. 
Contrary to the case of the SM, in the MSSM two Higgs doublets are required.
This results in five physical Higgs bosons instead of the single Higgs
boson in the SM.  These are the light and heavy $\cp$-even Higgs bosons, 
$h$ and $H$, the $\cp$-odd Higgs boson, $A$, and the charged Higgs bosons,
$H^\pm$.
The neutral SUSY partners of the (neutral) Higgs and electroweak gauge
bosons gives rise to the four neutralinos, $\neu{1,2,3,4}$.  The corresponding
charged SUSY partners are the charginos, $\cha{1,2}$.
The SUSY partners of the SM leptons and quarks are the scalar leptons
and quarks (sleptons, squarks), respectively.

The electroweak (EW) sector of the MSSM (the charginos,
neutralinos and scalar leptons) can account for a variety of experimental
data. The lightest SUSY particle (LSP), the lightest neutralino $\neu1$,
can explain the CDM relic abundance~\cite{Go1983,ElHaNaOlSr1984}, while
not being in conflict with negative DD results and the negative LHC searches. 
The requirement to give the full amount of DM relic density can be met if
the LSP is a bino or a mixed bino/wino state.
Furthermore, the EW sector of the MSSM can account for the persistent
$3-4\,\sig$ discrepancy of \gmin2. 
Recently in \citere{CHSold}, assuming that the LSP gives rise to the
full amount of DM relic density, upper limits on the
various masses of the EW SUSY sector were derived, while being in
agreement with all other experimental data%
\footnote{Other articles that investigated (part of) this interplay
are \citeres{Bharucha:2013epa,Fowlie:2013oua,Han:2013kza,Kowalska:2015zja,Choudhury:2016lku,Datta:2016ypd,Chakraborti:2017vxz,Hagiwara:2017lse,Yanagida:2020jzy,Yin:2016shg,Yanagida:2016kag,Chakraborti:2017dpu,Bagnaschi:2015eha,Datta:2018lup,Cox:2018qyi,Cox:2018vsv,Abdughani:2019wai,Endo:2020mqz,Pozzo:2018anw,Athron:2018vxy},
see \citere{CHSold} for a detailed description.}%
.
The upper limits strongly
depend on the deviation of the experimental result of \gmin2\ from its
SM prediction. Taking the current deviation
of \gmin2~\cite{Keshavarzi:2019abf,Davier:2019can}, limits of roughly
$\sim 600 \gev$ were set on the mass of the LSP and the next-to-LSP
(NLSP). Assuming that the new result of the
Run~1 of the ``MUON G-2'' collaboration at Fermilab yields a precision
comparable to the existing experimental result with the same central
value, yielded a reduction of these upper limits of roughly
$\sim 100 \gev$. 

In this paper we perform an analysis similar to~\cite{CHSold}, but under
the assumption that the relic DM only gives an upper bound, which opens
up the possibility of wino and higgsino DM. We analyze the four cases
of bino-dominated, bino/wino, (nearly pure) wino and higgsino LSP for the current
deviation of \gmin2, as well as for the assumption of improved \gmin2\
bounds from the combination of existing experimental data with the
Run\,1 data of the ``MUON G-2'' experiment.
In all cases we require the agreement with all other relevant existing
data, such as the DD bounds and the EW searches at the LHC. The derived
upper limits on the EW masses are discussed in the context of
the upcoming searches at the HL-LHC as well as at possible future
$e^+e^-$~colliders, such as the ILC~\cite{ILC-TDR,LCreport} or
CLIC~\cite{CLIC,LCreport}.


\section {The electroweak sector of the MSSM}
\label{sec:model}

In our notation for the MSSM we follow exactly \citere{CHSold}. Here we
retrict ourselves to a very short introduction of the relevant parameters and
symbols of the EW sector of the MSSM, consisting of charginos,
neutralinos and scalar leptons. The scalar quark sector is assumed to
be heavy and not to play a relevant role in our analysis. Throughout this
paper we also assume that all parameters are real, i.e.\ the absence of
$\CP$-violation.

The masses and mixings of the neutralinos are determined (besides SM
parameters)) by $U(1)_Y$ and $SU(2)_L$ 
gaugino masses $M_1$ and $M_2$, the Higgs mixing parameter $\mu$
and $\tb$, the ratio of the two
vacuum expectation values (vevs) of the two Higgs doublets of MSSM,
$\tb = v_2/v_1$.
After diagonalization, the four eigenvalues
of the matrix give the four neutralino masses
$\mneu1 < \mneu2 < \mneu3 <\mneu4$.
The masses and mixings of the charginos are determined (besides SM
parameters) by $M_2$, $\mu$ and $\tb$. 
Diagonalizing the mass matrix two chargino-mass
eigenvalues $\mcha1 < \mcha2$ can be obtained.

For the sleptons, as in \citere{CHSold}, we choose common soft
SUSY-breaking parameters for all three generations.
The charged slepton mass matrix are determined (besides SM parameters) by
the diagonal soft SUSY-breaking parameters $\mL^2$ and $\mR^2$ and the
trilinear coupling $A_l$ ($l = e, \mu, \tau$), where the latter are
taken to be zero. Mixing between the ``left-handed'' and
``right-handed'' sleptons is only relevant for scalar taus, where the
off-diagonal entry in the mass matrix is given by $-m_\tau \mu \tb$.
Thus, for the first two generations, the mass eigenvalues can be approximated
as $\msl1 \simeq \mL, \msl2 \simeq \mR$.
In general we follow the convention that $\Sl_1$ ($\Sl_2$) has the
large ``left-handed'' (``right-handed'') component.
Besides the symbols equal for all three generations, we also
explicitly use the scalar electron, muon and tau masses,
$\mse{1,2}$, $\msmu{1,2}$ and $\mstau{1,2}$.
The sneutrino and slepton masses are connected by the usual SU(2) relation.

Overall, the EW sector at the tree level
can be described with the help of six parameters: $M_1,M_2,\mu, \tb, \mL, \mR$.
Throughout our analysis we neglect $\cp$-violation and
assume $\mu, M_1, M_2 > 0 $. In \citere{CHSold} it was shown that
choosing these parameters positive covers the relevant parameter space
once the \gmin2\ results are taken into account
(see, however, the discussion in \refse{sec:conclusion}).


\medskip
Following the stronger experimental limits from the
LHC~\cite{ATLAS-SUSY,CMS-SUSY},
we assume that the colored sector of the MSSM is sufficiently heavier
than the EW sector, and does not play a role in this analysis. For the
Higgs-boson sector we assume that the radiative corrections to the light
$\cp$-even Higgs boson (largely originating from the top/stop sector)
yield a value in agreement with the experimental data, $\Mh \sim 125 \gev$.
This naturally yields stop masses in the TeV
range~\cite{Bagnaschi:2017tru,Slavich:2020zjv}, in agreement 
with the above assumption. Concerning the Higgs-boson mass scale, as
given by the $\cp$-odd Higgs-boson mass, $\MA$, we employ the existing
experimental bounds from the LHC. In the combination with other data,
this results in a mostly non-relevant impact of the heavy Higgs
bosons on our analysis, as will be discussed below.


\section {Relevant constraints}
\label{sec:constraints}


The experimental result for 
$\amu := \gmin2/2$ is dominated by the measurements made at the
Brookhaven National Laboratory (BNL)~\cite{Bennett:2006fi},
resulting in a world average of~\cite{PDG2018}
\begin{align}
\amu^{\rm exp} &= 11 659 209.1 (5.4) (3.3) \times 10^{-10}~,
\label{gmt-exp}
\end{align}
where the first uncertainty is statistical and the second systematic.
The SM prediction of \amu\ is given by~\cite{Aoyama:2020ynm}
(based on \citeres{Aoyama:2012wk,Aoyama:2019ryr,Czarnecki:2002nt,Gnendiger:2013pva,Davier:2017zfy,Keshavarzi:2018mgv,Colangelo:2018mtw,Hoferichter:2019mqg,Davier:2019can,Keshavarzi:2019abf,Kurz:2014wya,Melnikov:2003xd,Masjuan:2017tvw,Colangelo:2017fiz,Hoferichter:2018kwz,Gerardin:2019vio,Bijnens:2019ghy,Colangelo:2019uex,Blum:2019ugy,Colangelo:2014qya}
)%
\footnote{
In \citere{CHSold} a slightly different value was used, with a
negligible effect on the results.
}%
, 
\begin{align}
\amu^{\rm SM} &= (11 659 181.0 \pm 4.3) \times 10^{-10}~.
\label{gmt-sm}
\end{align}
Comparing this with the current experimental measurement in \refeq{gmt-exp}
results in a deviation of
\begin{align}
\Delta\amu^{\rm old} &= (28.1 \pm 7.6) \times 10^{-10}~, 
\label{gmt-diff}
\end{align}
corresponding to a $3.7\,\sig$ discrepancy.
This ``current'' result will be used below with a hard cut at 
$2\,\sig$ uncertainty.

Efforts to improve the experimental result at Fermilab by the
``MUON G-2'' collaboration~\cite{Grange:2015fou}
and at J-PARC~\cite{Mibe:2010zz} aim to reduce the experimental
uncertainty by a factor of four compared to the BNL measurement. 
For the second step in our analysis we consider the upcoming Run~1
result from the Fermilab experiment~\cite{gmt-new}.
The Run~1 data is expected to have roughly the same
experimental uncertainty as the current result in \refeq{gmt-exp}. 
We furthermore assume that the Run~1 data yields
the same central value as the current result. Consequently,
we anticipate that the experimental uncertainty 
shrinks by $1/\sqrt{2}$, yielding a future value of
\begin{align}
\De\amu^{\rm fut} & = (28.1 \pm 6.2) \times 10^{-10}~,
\label{gmt-fut}
\end{align}
corresponding to a $4.5\,\sig$ discrepancy.
Thus, the combination of Run~1 data with the existing
experimental \gmin2\ data has the potential to (nearly) establish the
``discovery'' of BSM physics.
This ``anticipated future'' result will be used below with a hard cut at 
$2\,\sig$ uncertainty.

\medskip
Recently a new lattice calculation for the leading order hadronic
vacuuum polarization (LO HVP) contribution to
$\amu^{\rm SM}$~\cite{Borsanyi:2020mff} has been reported, which,
however, was not used in the new theory world
average, \refeq{gmt-sm}~\cite{Aoyama:2020ynm}. Consequently, we also do
not take this result into account, see also the discussions in
\citere{CHSold,Lehner:2020crt,Borsanyi:2020mff,Crivellin:2020zul,Keshavarzi:2020bfy,deRafael:2020uif}. 
On the other hand, we are also
aware that our conclusions would change substantially if the result
presented in \cite{Borsanyi:2020mff} turned out to be correct.

\medskip
In the MSSM the main contribution to \gmin2\ at the one-loop level comes from
diagrams involving $\cha1-\Sn$ and $\neu1-\tilde \mu$ loops. In the
case of a bino-dominated LSP the contributions are approximated
as~\cite{Moroi:1995yh,Martin:2001st,Badziak:2019gaf} 
\label{subsec:mssmgmin2}
\begin{eqnarray}
\label{amuchar}
a^{\tilde \chi^{\pm}-\Sn_{\mu}}_{\mu} &\approx&
\frac{\alpha \, m^2_\mu \, \mu\,M_{2} \tb}
{4\pi \sin^2\theta_W \, m_{\tilde{\nu}_{\mu}}^{2}}
\left( \frac{f_{\chi^{\pm}}(M_{2}^2/m_{\tilde{\nu}_{\mu}}^2)
-f_{\chi^{\pm}}(\mu^2/m_{\tilde{\nu}_{\mu}}^2)}{M_2^2-\mu^2} \right) \, ,
\\
\label{amuslep}
a^{\tilde \chi^0 -\tilde \mu}_{\mu} &\approx&
\frac{\alpha \, m^2_\mu \, \,M_{1}(\mu \tb-A_\mu)}
{4\pi \cos^2\theta_W \, (m_{\tilde{\mu}_R}^2 - m_{\tilde{\mu}_L}^2)}
\left(\frac{f_{\chi^0}(M^2_1/m_{\tilde{\mu}_R}^2)}{m_{\tilde{\mu}_R}^2}
- \frac{f_{\chi^0}(M^2_1/m_{\tilde{\mu}_L}^2)}{m_{\tilde{\mu}_L}^2}\right)\,,
\end{eqnarray}
where the loop functions $f$ are as given in Ref.~\cite{Badziak:2019gaf}.
In our analysis MSSM contribution to \gmin2\
up to two-loop order is calculated using {\tt
GM2Calc}~\cite{Athron:2015rva}, implementing two-loop corrections
from \cite{vonWeitershausen:2010zr,Fargnoli:2013zia,Bach:2015doa}
(see also \cite{Heinemeyer:2003dq,Heinemeyer:2004yq}).
This code also works numerically reliable for the cases of (very)
compressed spectra, such as for wino and higgsino DM.


\subsection*{Other constraints}

All other experimental constraints are taken into account exactly as
in \citere{CHSold}. These comprise

\begin{itemize}

\item Vacuum stability constraints:\\
All points are check to possess a stable and correct EW vacuum, e.g.\
avoiding charge and color breaking minima. This check is performed with
the public code {\tt Evade}~\cite{Hollik:2018wrr,Robens:2019kga}.

\item Constraints from the LHC:\\
All relevant EW SUSY searches are taken into account, mostly via
\CM~\cite{Drees:2013wra,Kim:2015wza, Dercks:2016npn}, where many
analysis had to be implemented newly~\cite{CHSold}.
As in \citere{CHSold}, the constraints coming from
"compressed spectra" searches~\cite{Aad:2019qnd},
corresponding to very low splittings between $\mcha1,\mneu2,\msl1$
and $\mneu1$ are applied directly on our parameter space.

In addition to the searches described in \citere{CHSold},
we take into account the latest constraints from the disappearing track
searches at the LHC~\cite{Aaboud:2017mpt,Sirunyan:2020pjd}.
These are particularly important for wino DM scenario
where the mass gap between $\chapm1$ and $\neu1$ can be $\sim$ a few
hundred $\mev$. The long-lived $\chapm1$ (lifetime $\sim \order{\mathrm{ns}}$)
decays into final states involving a $\neu1$ and a soft pion which can not be
reconstructed within the detector. Thus, the signal involves a charged
track from $\chapm1$ that produces hits only in the innermost layers
of the detector with no subsequent hits at larger radii.

\item
Dark matter relic density constraints:\\
We use the latest result from Planck~\cite{Planck}.
\begin{align}
\Omega_{\rm CDM} h^2 \; \le \; 0.122, 
\label{OmegaCDM}
\end{align}
As stressed above, we take the relic density as an {\it upper} limit
(evaluated from the central value plus $2\,\sig$.
The relic density in the MSSM is evaluated with
\MO~\cite{Belanger:2001fz,Belanger:2006is,Belanger:2007zz,Belanger:2013oya}.
An additional DM component could be, e.g., a SUSY
axion~\cite{Bae:2013bva},
which would then bring the total DM density into agreement with the
Planck measurement of $\Omega_{\rm CDM} h^2 = 0.120 \pm 0.001$~\cite{Planck}.

In the case of wino DM, because of the extremely small mass
splitting, the effect of ``Sommerfeld
enhancement''~\cite{Sommerfeld1931} can be very 
important. For wino DM providing the full amount of DM it shifts the
allowed range of $\mneu1$ from $\sim 2.0 \tev$ to about $\sim 2.9 \tev$. 
Since here we are interested in the case that the wino DM only gives a
fraction of the whole DM relic density, see \refeq{OmegaCDM}, we can
safely neglect the Sommerfeld enhancement. The upper limit on $\mneu1$
is given, as will be shown below, by the $\gmin2$ constraint, but not by
the DM relic density. Allowing higher masses here (as would be the case
if the Sommerfeld enhancement had been taken into account) could thus
not lead to a larger allowed parameter space. On the other hand, for a point
with a relic density fulfilling \refeq{OmegaCDM} the Sommerfeld
enhancement would only lower the ``true'' DM density, which still
fulfills \refeq{OmegaCDM}.

\item
Direct detection constraints of Dark matter:\\
We employ the constraint on the spin-independent
DM scattering cross-section $\ssi$ from
XENON1T~\cite{XENON} experiment, 
evaluating the theoretical prediction for $\ssi$ using
\MO~\cite{Belanger:2001fz,Belanger:2006is,Belanger:2007zz,Belanger:2013oya}.
A combination with other DD experiments would yield only very slightly
stronger limits, with a negligible impact on our results.
For parameter points with $\Omega_{\tilde \chi} h^2 \; \le \; 0.118$
($2\,\sig$ lower limit from Planck~\cite{Planck}), we scale the cross-section
with a factor of ($\Omega_{\tilde \chi} h^2$/0.118) 
to account for the fact
that $\neu1$ provides only a fraction of the total DM relic density of
the universe.
Here the effect of neglecting the Sommerfeld enhancement leads to a
more conservative allowed region of parameter space.

\end{itemize}


\section{Parameter scan and analysis flow}
\label{sec:paraana}

\subsection{Parameter scan}
\label{sec:scan}

We scan the relevant MSSM parameter space to obtain lower and {\it upper}
limits on the relevant neutralino, chargino and slepton masses.
In order to achieve a ``correct'' DM relic density, see \refeq{OmegaCDM},
by the lightest neutralino, $\neu1$, some mechanism such as a specific
co-annihilation or pole annihilation has to be active in the early
universe. At the same time $\mneu1$ must not be too high, such that the 
EW sector can provide the contribution required to bring the theory
prediction of $\amu$ into agreement with the experimental measurement,
see \refse{sec:constraints}. 
The combination of these two requirements yields the following
possibilities. (The cases present a certain choice of favored
possibilities, upon which one can expand, as will briefly discussed
in \refse{sec:conclusion}.)

\begin{description}
\item
{\bf (A) Higgsino DM}\\
This scenario is characterized by a small value of~$\mu$ (as favored,
e.g., by naturalness
arguments~\cite{Baer:2012up,Baer:2013gva,Baer:2016lpj,Baer:2018rhs,Bae:2019dgg,Baer:2020vad})%
\footnote{
A recent analysis in the higgsino DM scenario, requiring the LSP to
yield the full DM relic density, can be found in \citere{Delgado:2020url}.
}%
. Such a scenario is also naturally realized in Anomaly Mediation SUSY
breaking (see e.g.\ \citere{Bagnaschi:2016xfg} and references therein).
We scan the following parameters: 
\begin{align}
  100 \gev \leq \mu \leq 1.2 \tev \;,
  \quad 1.1 \mu \leq M_1 \leq 10 \mu\;, \notag \\
  \quad 1.1  \mu \leq M_2 \leq 10 \mu, \;
  \quad 5 \leq \tb \leq 60, \; \notag\\
  \quad 100 \gev \leq \mL, \mR \leq 2  \tev~.
\label{wino-dm}
\end{align}

\item
{\bf (B) Wino DM}\\
This scenario is characterized by a small value of $M_2$.
Such a scenario is also naturally realized in Anomaly Mediation SUSY
breaking (see e.g.\ \citere{Bagnaschi:2016xfg} and references therein).
We scan the following parameters: 
\begin{align}
  100 \gev \leq M_2 \leq 1.5 \tev \;,
  \quad 1.1 M_2 \leq M_1 \leq 10 M_2\;, \notag \\
  \quad 1.1 M_2 \leq \mu \leq 10 M_2, \;
  \quad 5 \leq \tb \leq 60, \; \notag\\
  \quad 100 \gev \leq \mL, \mR \leq 2 \tev~.
\label{wino-dm}
\end{align}
The choice of $M_2 \ll M_1, \mu$ leads (at tree-level) to a very
degenerate spectrum with $\mcha1 - \mneu1 = \order{1 {\rm\,eV}}$.
However, this spectrum does not correspond to the on-shell (OS)
masses of all six charginos and neutralinos. Since only three (soft
SUSY-breaking) mass parameters are available ($M_1$, $M_2$ and $\mu$),
only three out of the six masses can be renormalized OS. The
(one-loop) shifts for the three remaining masses are obtained via the
CCN$_i$ renormalization scheme~\cite{Fritzsche:2013fta} with
$i \in \{2, 3, 4\}$.
In a CCN$_i$ scheme the $\cha1$, $\cha2$ and $\neu{i}$ are chosen
OS. This automatically yields a good renormalization for $M_2$ and
$\mu$. The $\neu{i}$ has to be chosen, parameter point by parameter
point, such that also $M_1$ is renormalized well (which excludes the
CCN$_1$ for $M_2 \ll M_1, \mu$). For each point we choose the
$\neu{i}$ to be renormalized OS such that the maximum shift of all three
shifted neutralino 
masses is minimized. We have explicitly checked for each scanned
point that the such chosen CCN$_i$ indeed yields
reasonably small shifts for three shifted neutralino masses, in
particular for $\mneu1$.%
\footnote{We thank C.~Schappacher for evaluating the mass shift for our
wino DM points (following \citere{Heinemeyer:2017izw}).}%
~Only this transition to OS masses yields a mass splitting between
$\mcha1$ and $\mneu1$ that allows then for the decay
$\chapm1 \to \neu1 \pi^\pm$.

\item
{\bf (C) Mixed bino/wino DM}\\
This scenario has been analyzed in \citere{CHSold}. It can in
principle be realized
in three different versions corresponding to the coannihilation
mechanism (see \citere{CHSold} for a detailed discussion). However,
a larger wino component is found only for $\chapm1$-coannihilation.
The scan parameters are chosen as, 
\begin{align}
  100 \gev \leq M_1 \leq 1 \tev \;,
  \quad M_1 \leq M_2 \leq 1.1 M_1\;, \notag \\
  \quad 1.1 M_1 \leq \mu \leq 10 M_1, \;
  \quad 5 \leq \tb \leq 60, \; \notag\\
  \quad 100 \gev \leq \mL \leq 1.5 \tev, \; 
  \quad \mR = \mL~.
\label{cha-coann}
\end{align}
Here we choose one soft SUSY-breaking parameter for all sleptons
together. While this choice should not have a relevant effect in the
$\cha1$-coannihilation case, this have an impact in the next case.
In our scans we will see that the chosen lower and upper limits are not
reached by the points that meet all the experimental constraints. This
ensures that the chosen intervals indeed cover all the relevant
parameter space. 

\item
{\bf (D) Bino DM}\\
This scenario covers the coannihilation with sleptons, and has also been
analyzed in \citere{CHSold}. In this scenario ``accidentally'' the wino
component of the $\neu1$ can be non-negligible. However, this is not a
distinctive feature of this scenario.
We cover the two distinct cases that either the SU(2)
doublet sleptons, or the singlet sleptons are close in mass to the LSP.\\
{\bf (D1)} Case-L: SU(2) doublet
\begin{align}
  100 \gev \leq M_1 \leq 1 \tev \;,
  \quad M_1 \leq M_2 \leq 10 M_1 \;, \notag\\
  \quad 1.1 M_1 \leq \mu \leq 10 M_1, \;
  \quad 5 \leq \tb \leq 60, \; \notag\\
  \quad M_1 \gev \leq \mL \leq 1.2 M_1, 
  \quad M_1 \leq \mR \leq 10 M_1~. 
\label{slep-coann-doublet}
\end{align}

{\bf (D2)} Case-R: SU(2) singlet
\begin{align}
  100 \gev \leq M_1 \leq 1 \tev \;,
  \quad M_1 \leq M_2 \leq 10 M_1 \;, \notag \\
  \quad 1.1 M_1 \leq \mu \leq 10 M_1, \;
  \quad 5 \leq \tb \leq 60, \; \notag\\
  \quad M_1 \gev \leq \mR \leq 1.2 M_1,\; 
  \quad M_1 \leq \mL \leq 10 M_1~.
\label{slep-coann-singlet}
\end{align}
\end{description}
In all scans we choose flat priors of the parameter space and
generate \order{10^7} points.

The mass parameters of the colored sector have been set to high values,
such that 
the resulting SUSY particle masses are outside the reach of the LHC, the
light $\cp$-even Higgs-boson is in agreement with the LHC measurements
(see, e.g., \citeres{Bagnaschi:2017tru,Slavich:2020zjv}),
where the concrete values are not relevant for our analysis. $\MA$ has
also been set to be above the TeV scale. Consequently, we
do not include explicitly the possibility of $A$-pole annihilation,
with $\MA \sim 2 \mneu1$. As we will discuss below the combination of
direct heavy Higgs-boson searches with the other experimental
requirements constrain this possibility substantially (see,
however, also \refse{sec:future}).
Similarly, we do not consider $h$-~or $Z$-pole annihilation, as such a
light neutralino sector likely overshoots the \gmin2\ contribution
(see, however, the discussion in \refse{sec:future}).


\subsection{Analysis flow}
\label{sec:flow}

The data sample is generated by scanning randomly over the input parameter
range mentioned above, using a flat prior for all parameters.
We use {\tt SuSpect}~\cite{Djouadi:2002ze}
as spectrum and SLHA file generator, which yields \DRbar\ values
for the chargino and neutralino masses. While in most of the cases the
difference between \DRbar\ and on-shell (OS) masses is not
phenomenologically relevant for our analysis, this is different for wino DM.
Because of the very small mass difference of the \DRbar\ values for
$\mneu1$ and $\mcha1$ we take into account the transition to OS
masses, as discussed above. 
In the next step the points are required
to satisfy the $\chapm1$ mass limit from LEP~\cite{lepsusy}. 
The SLHA output files
from {\tt SuSpect} are then passed as input to {\tt GM2Calc} and \MO~for
the calculation of \gmin2 and the DM observables, respectively. The parameter
points that satisfy the current \gmin2\ constraint, \refeq{gmt-diff}, the DM
relic density, \refeq{OmegaCDM}, the direct detection constraints and
the vacuum stability constraints as checked with {\tt Evade} are
then taken to the final 
step to be checked against the latest LHC constraints implemented in \CM.
~The branching ratios of the relevant SUSY
particles are computed using {\tt SDECAY}~\cite{Muhlleitner:2003vg}
and given as input to \CM.



\section{Results}
\label{sec:results}


\subsection{Higgsino DM}
\label{sec:higgsino}

We start our discussion with the case of higgsino DM, as discussed
in \refse{sec:scan}. We follow the analysis flow as described
in \refse{sec:flow}, where the vacuum stability constraints did not
affect any of the LHC allowed points. We also remark that we have
checked that the (possible) one-loop corrections to the
chargino/neutralino masses do not change the results in a relevant way
and were thus omitted (see, however, \refse{sec:wino}).
In the following we denote the
points surviving certain constraints with different colors:
\begin{itemize}
\item grey (round): all scan points (i.e.\ points excluded by \gmin2).
\item green (round): all points that are in agreement with \gmin2, taking
into account the current or anticipated future limits,
see \refeqs{gmt-diff} and (\ref{gmt-fut}), respectively,
(i.e.\ points excluded by the DM relic density).
\item blue (triangle): points that additionally obey the upper limit of
the DM relic density, see \refeq{OmegaCDM} (i.e.\ points excluded
by the DD constraints). In some plots below all points that pass
the \gmin2\ constraint are also in agreement with the DM relic density
constraint, resulting in only blue (but no green) points to be visible.
\item cyan (diamond): points that additionally pass the DD constraints,
see \refse{sec:constraints} (i.e.\ points excluded by the LHC
constraints). 
\item red (star):  points that additionally pass the LHC constraints,
see \refse{sec:constraints} (i.e.\ points that pass all constraints).
\end{itemize}

\begin{figure}[htb!]
\centering
\begin{subfigure}[b]{0.48\linewidth}
\centering\includegraphics[width=\textwidth]{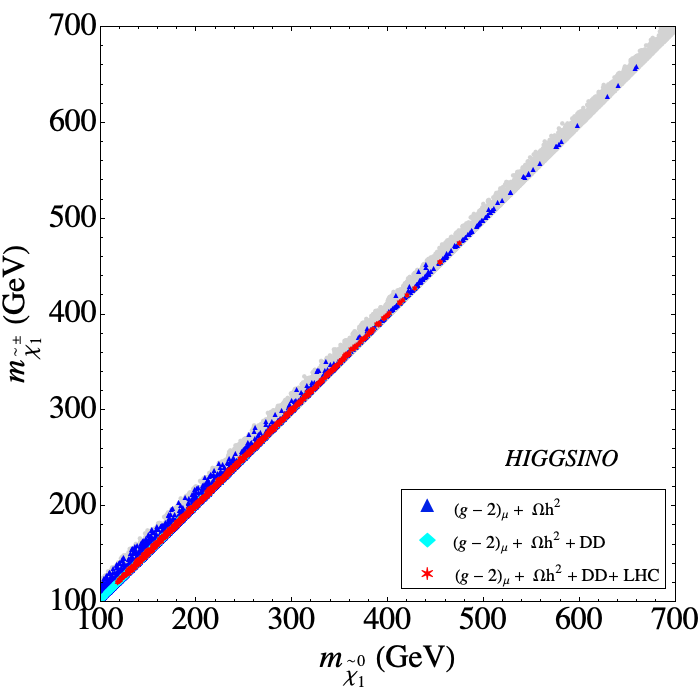}
	\caption{}
	\label{}
\end{subfigure}
~
\begin{subfigure}[b]{0.48\linewidth}
\centering\includegraphics[width=\textwidth]{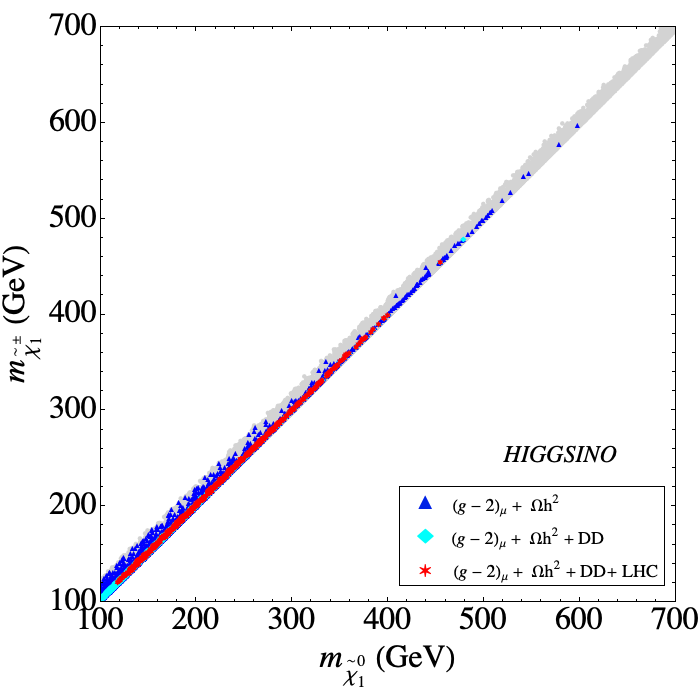}
	\caption{}
	\label{}
\end{subfigure}
\caption{The results of our parameter scan in the $\mneu1$--$\mcha1$ plane
for the higgsino DM scenario for current (left) and anticipated future
limits (right) from \gmin2. For the color coding: see text.
}
\label{mn1-mc1-higgsino}
\end{figure}

In \reffi{mn1-mc1-higgsino} we show our results in the $\mneu1$--$\mcha1$
plane for the current (left) and future (right) \gmin2\ constraint,
see \refeqs{gmt-diff} and (\ref{gmt-fut}), respectively. By definition
the points are clustered in the diagonal of the plane, and
$\mneu2 \approx \mcha1$.
Starting with the \gmin2\ constraint (green points), they all pass also
the relic density constraint (dark blue) and thus are visible as
dark blue points in
the plot. Overall, one can observe a clear upper limits from \gmin2\ of
about $660 \gev$ for the current limits and about $600 \gev$ from the
anticipated future accuracy. As mentioned above, the DM relic density
constraint does not yield changes in the allowed parameter space.
Applying the DD limits, on the other hand,
forces the points to have even smaller mass differences between $\neu1$
and $\cha1$. It has also an important impact on the upper limit, which
is reduced to about $500~(480) \gev$ for the current (future) \gmin2\
bounds. Applying the LHC
constraints, corresponding to the ``surviving'' red points (stars), does
not yield a further reduction from above for the current \gmin2\
constraint, whereas for the anticipated future accuracy yields a
reduction of $\sim 30 \gev$. The LHC constraints (see also the
discussion below) also cut always (as anticipated) points in the lower
mass range, resulting in a lower limit of $\sim 120 \gev$ for
$\mneu1 \approx \mcha1$. 

\begin{figure}[htb!]
\vspace{2em}
\centering
\begin{subfigure}[b]{0.48\linewidth}
\centering\includegraphics[width=\textwidth]{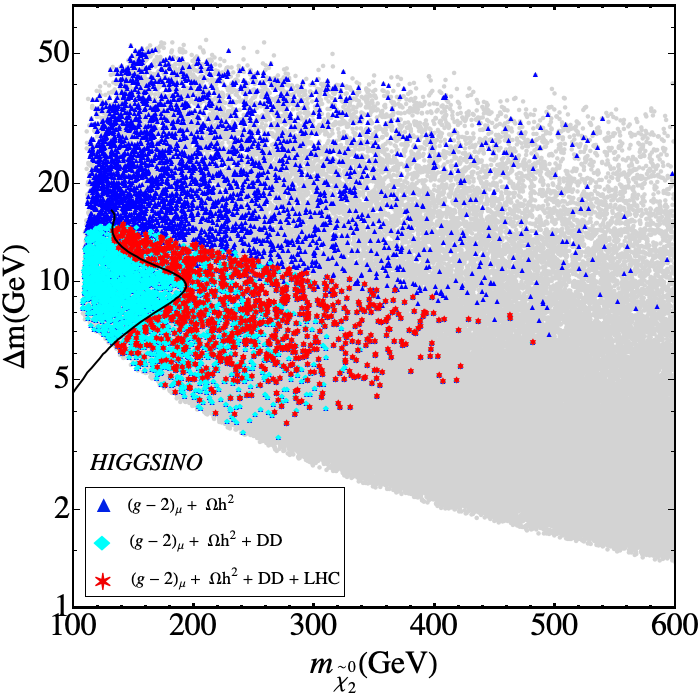}
	\caption{}
	\label{}
\end{subfigure}
~
\begin{subfigure}[b]{0.48\linewidth}
\centering\includegraphics[width=\textwidth]{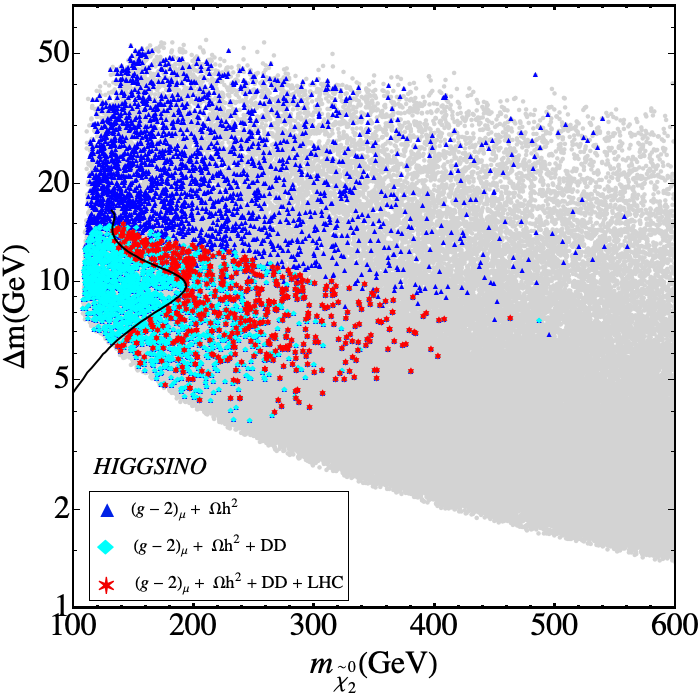}
	\caption{}
	\label{}
\end{subfigure}
\caption{The results of our parameter scan in the
$\mneu2$--$\Delta m(= \mneu2-\mneu1)$ plane  
for the higgsino DM scenario for current (left) and anticipated future
limits (right) from \gmin2. For the color coding: see text.
}
\label{mn2-delm-higgsino}
\end{figure}

The LHC constraint which is most effective in this parameter plane is the
one designed for compressed spectra, as demonstrated
in \reffi{mn2-delm-higgsino}, showing our scan points in the
$\mneu1$--$\De m(= \mneu2 - \mneu1)$ parameter plane. 
The color coding is as in \reffi{mn1-mc1-higgsino}. 
The black line indicates the bound from compressed
spectra searches~\cite{Aad:2019qnd}. 
One can clearly observe that the compressed spectra bound sits
exactly in the preferred higgsino DM region, i.e.\ at
$\De m(= \mneu2 - \mneu1) = \order{10 \gev}$.
Other LHC constraint that is effective in this case is the
bound from slepton pair production leading to dilepton and $\met$
in the final state~\cite{Aad:2019vnb}, as will be discussed in
more detail below. The bounds from the disappearing track
searches~\cite{Sirunyan:2020pjd} turn out to be 
ineffective in this case because of the very short lifetime
of the $\chapm1$~\cite{Fukuda:2017jmk}.
From \reffis{mn1-mc1-higgsino}, \ref{mn2-delm-higgsino} one can conclude
that the experimental data set an upper as well as a lower bound,
yielding a clear search target for the upcoming LHC runs, and in
particular for future $e^+e^-$ colliders, as will be discussed
in \refse{sec:future}. In particular, this collider target gets
(potentially) sharpened by the improvement in the \gmin2\
measurements.

\begin{figure}[htb!]
\vspace{2em}
\centering
\begin{subfigure}[b]{0.48\linewidth}
\centering\includegraphics[width=\textwidth]{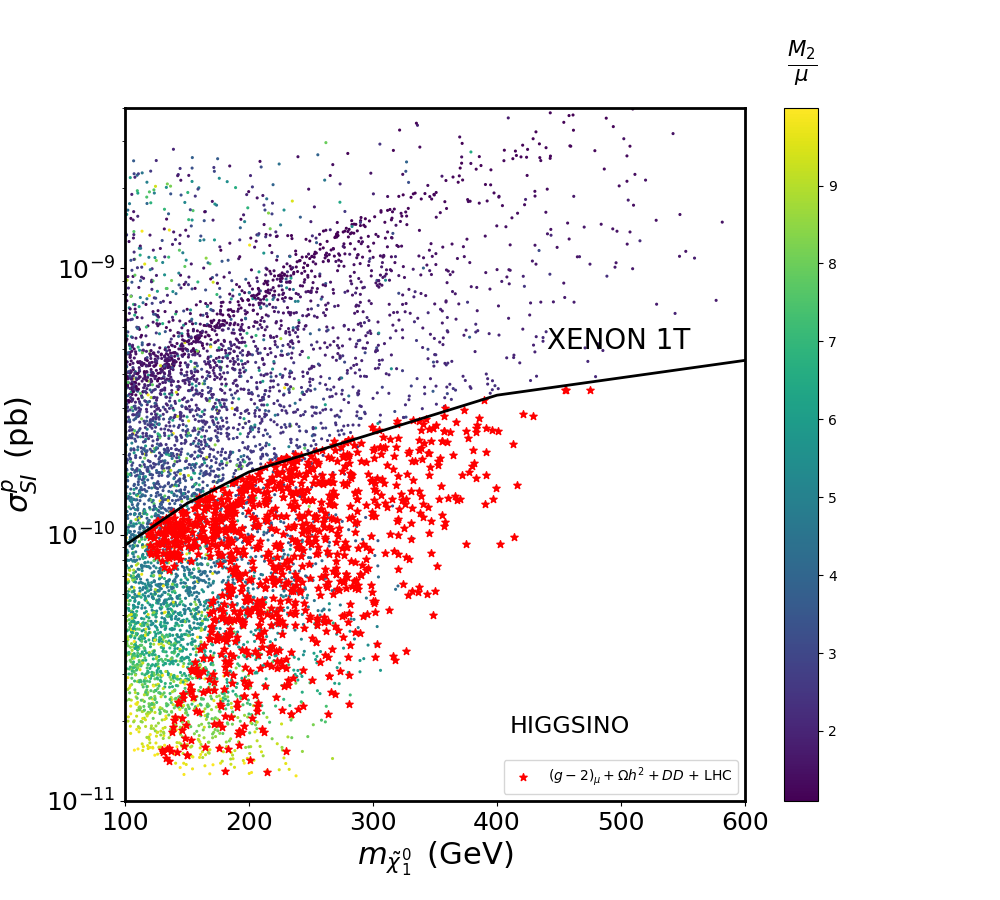}
	\caption{}
	\label{}
\end{subfigure}
~
\begin{subfigure}[b]{0.48\linewidth}
\centering\includegraphics[width=\textwidth]{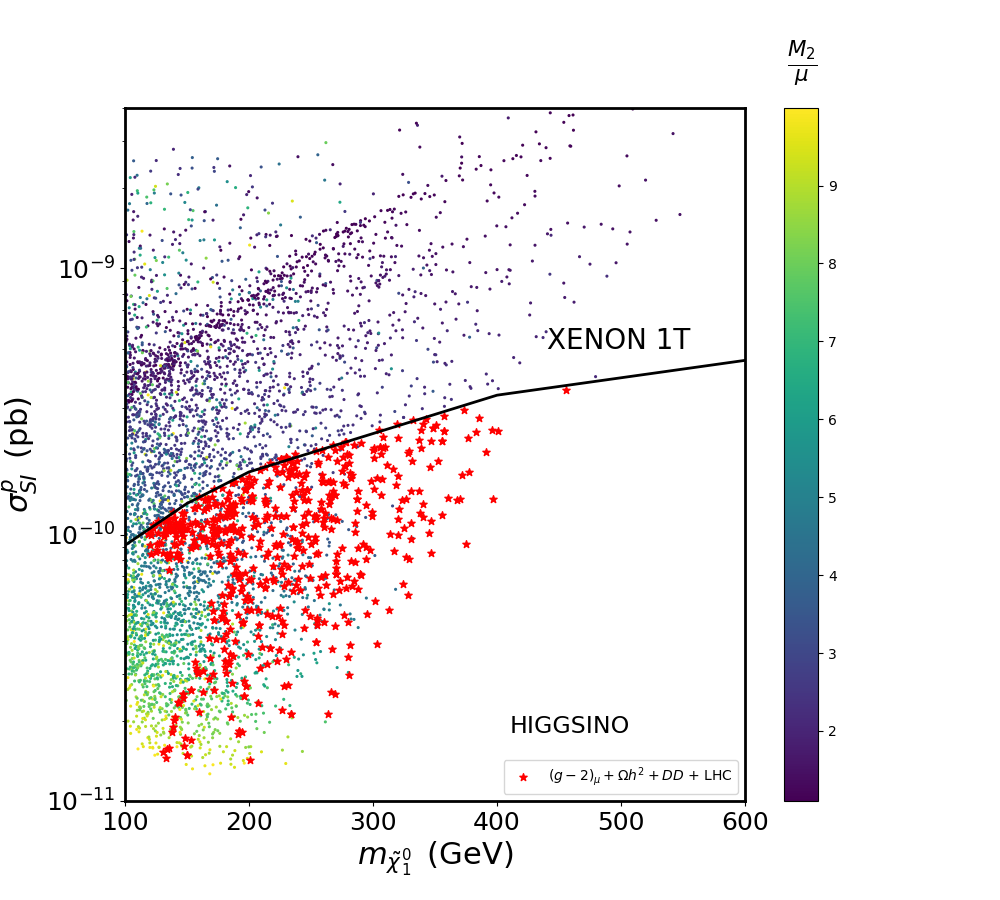}
	\caption{}
	\label{}
\end{subfigure}
\caption{Scan results in the $\mneu1$--$\ssi$ plane for higgsino DM
scenario for current (left)
and anticipated future limits (right) from \gmin2. The color coding of
the points denotes $M_2/\mu$ and the black line indicates the DD
limits (see text). In red we show the points fulfilling the \gmin2\ ,
relic density, DD and additionally the LHC constraints.
}
\label{mneu1-ssi-higgsino}
\end{figure}

The impact of the DD experiments is demonstrated 
in \reffi{mneu1-ssi-higgsino}. We show the $\mneu1$--$\ssi$ plane for
current (left) and anticipated future limits (right) from \gmin2. 
The color coding of the points (from yellow to dark blue)
denotes $M_2/\mu$, whereas in red we show the points fulfilling
\gmin2\ , relic density, DD and the LHC constraints.
The black line indicates the current DD limits, here
taken for sake of simplicity from XENON1T~\cite{XENON}, as discussed
in \refse{sec:constraints}. It can be seen that a slight downward shift of
this limit, e.g.\ due to additional DD experimental limits from
LUX~\cite{LUX} or PANDAX~\cite{PANDAX}, would not
change our results in a strong way, but only slightly reduce the upper
limit on $\mneu1$.
The scanned parameter space extends from large $\ssi$
values, given for the smallest scanned $M_2/\mu$ values to the
smallest ones, reached for the largest scanned $M_2/\mu$, i.e.\
the $\ssi$ constraints are particularly strong for small $M_2/\mu$,
which can be understood as
follows. The most important contribution to DM scattering comes
from the exchange of a light $\cp$-even Higgs
boson in the t-channel. The corresponding $h\neu1\neu1$ coupling
at tree level is given by~\cite{Hisano:2004pv}
\begin{eqnarray}
c_{h\neu1\neu1}&\simeq&
- \frac12 (1 + \sin2\beta)
\LP \tan^2\theta_\mathrm{w} \frac{\MW}{M_1-\mu}
+ \frac{\MW}{M_2-\mu} \RP\,,
\label{ddhiggsino}
\end{eqnarray}
where we have assumed $\mu > 0$. Thus, the coupling
becomes large for $\mu \sim M_2$ or $\mu \sim M_1$.
Therefore, the XENON1T DD bound pushes the allowed parameter space into
the almost pure higgsino-LSP region, with negligible bino and wino component.
The impact of the compressed spectra searches is visible in the lower
$\mneu1$ region.
Given both CDM constraints and the LHC
constraints, shown in red, the smallest $M_2/\mu$ value we find
is~2.2 for both current and
anticipated future \gmin2\ bound.
This result depends mildly on the
assumed \gmin2\ constraint, as this cuts away the largest $\mneu1$
values. All of the points will be conclusively probed by the future DD
experiment XENONnT~\cite{Aprile:2020vtw}.

\begin{figure}[htb!]
\centering
\begin{subfigure}[b]{0.48\linewidth}
\centering\includegraphics[width=\textwidth]{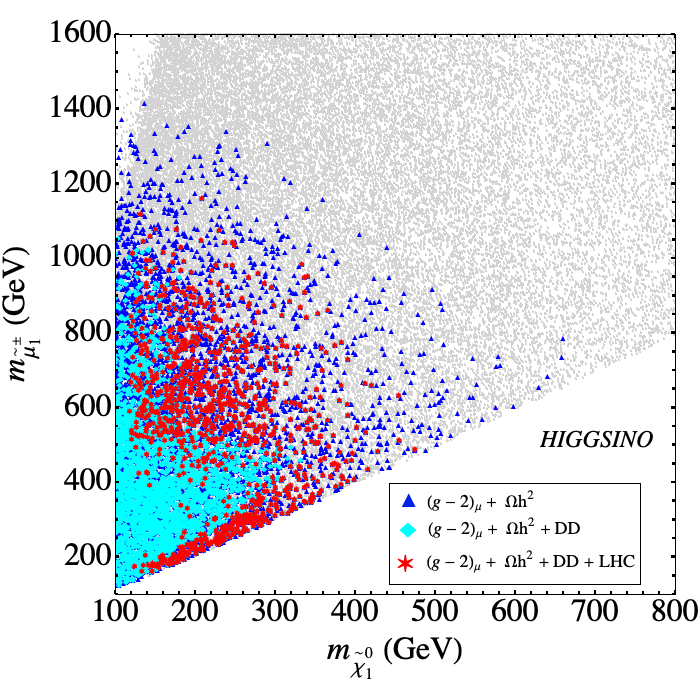}
\caption{}
\label{}
\end{subfigure}
~
\begin{subfigure}[b]{0.48\linewidth}
\centering\includegraphics[width=\textwidth]{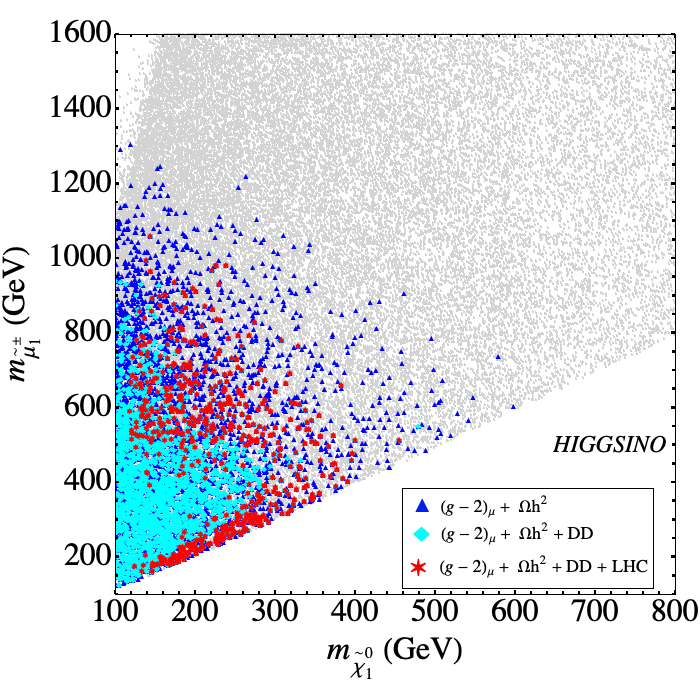}
\caption{}
\label{}
\end{subfigure}
\caption{The results of our parameter scan in the
$\mneu1$--$\msl1$ plane
for the higgsino DM scenario for current (left) and anticipated future
limits (right) from \gmin2.
The color coding is as in \protect\reffi{mn1-mc1-higgsino}.
}
\label{mneu1-mslep-higgsino}
\end{figure}

The distribution of $\msl1$ (where it should be kept in
mind that we have chosen the same masses for all three generations,
see \refse{sec:model}) is presented in the $\mneu1$--$\msl1$ plane
in \reffi{mneu1-mslep-higgsino}, with the same color coding as
in \reffi{mn1-mc1-higgsino}.
The \gmin2\ constraint places important constraints in this mass plane,
since both types of masses enter into the contributing SUSY diagrams,
see \refse{sec:constraints}. 
The constraint is satisfied in a roughly triangular region with its tip
around $(\mneu1, \msl1) \sim (650 \gev, 700 \gev)$ in the case of
current \gmin2\ constraints, and around $\sim (600 \gev, 600 \gev)$ in
the case of the anticipated future limits, i.e.\ the impact of the
anticipated improved limits is clearly visible as an {\it upper} limit
for both masses. 
Since no specific other requirement is placed on the slepton sector in the
higgsino DM case the slepton masses are distributed over the
\gmin2\ allowed region.
The DM relic density constraint, as discussed above, does not yield any
further bounds on the allowed parameter space. The inclusion of the DM DD
bounds, as visible by the cyan and red points, only cuts away the very
largest slepton masses (for a given LSP mass). 

The LHC constraints cut out all points with $\mneu1 \lsim 125 \gev$, as
well as a triangular region with the tip around
$(\mneu1, \msl1) \sim (340 \gev, 450 \gev)$. The first ``cut'' is due to the
searches for compressed spectra. The second cut is mostly a result
of the constraint coming from slepton pair production searches leading
to dilepton and $\met$ in the final state~\cite{Aad:2019vnb}. The bound
obtained by recasting the experimental search in \CM\ is substantially weaker
than the original limit from ATLAS. That limit is obtained for
a ''simplified model'' with $\br(\sle1, \sle2 \rightarrow l \neu1) = 100 \%$,
an assumption which is not stritly valid in our parameter space.
The small mass gap among $\neu1, \neu2$ and $\chapm1$ allows
significant $\br$ of the sleptons to final states involving $\chapm1$
and $\neu2$. This reduces the number of signal leptons and hence weakens
the exclusion limit.
Overall we can place an upper limit on the light slepton
mass of about $\sim 1200 \gev$ and $1050 \gev$ for the current and the
anticipated future accuracy of \gmin2, respectively. Since larger
values of slepton masses are reached for lower values of $\mneu1$, the
impact of \gmin2\ is relatively weaker than in the case of
chargino/neutralino masses.

\begin{figure}[htb!]
\centering
\begin{subfigure}[b]{0.48\linewidth}
\centering\includegraphics[width=\textwidth]{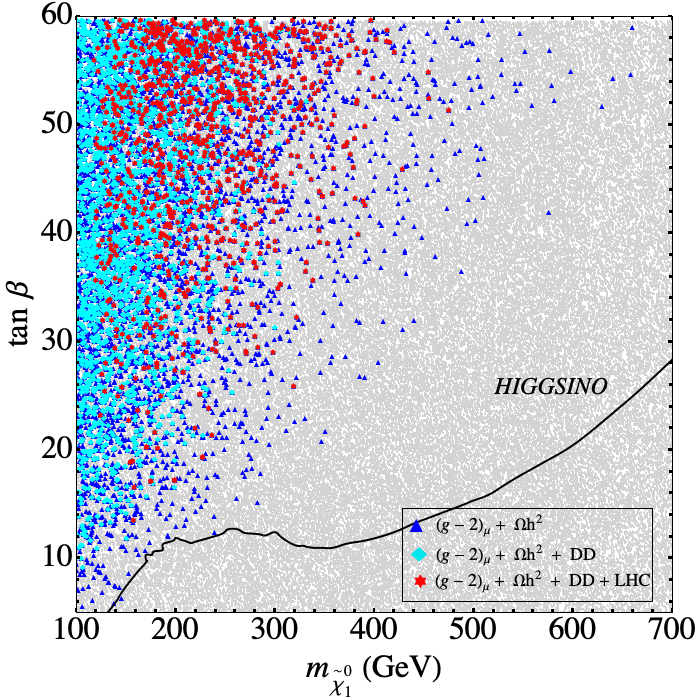}
	\caption{}
	\label{}
\end{subfigure}
~
\begin{subfigure}[b]{0.48\linewidth}
\centering\includegraphics[width=\textwidth]{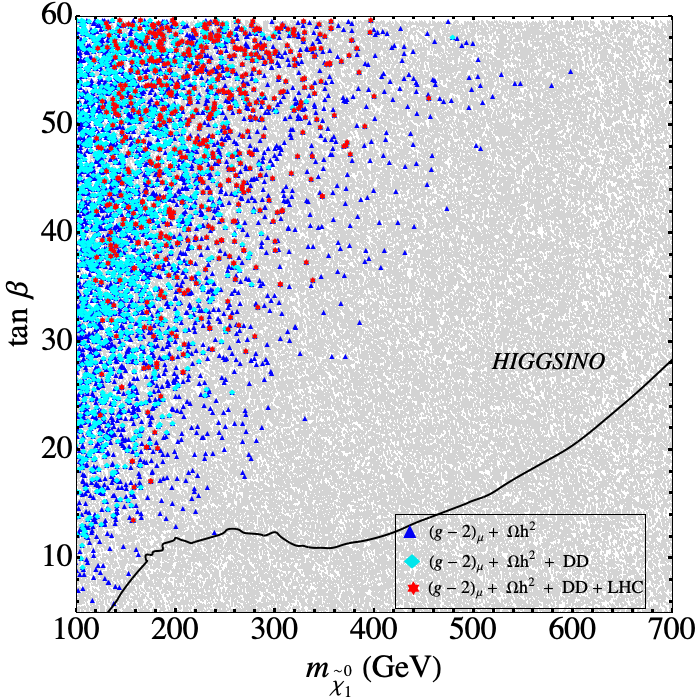}
	\caption{}
	\label{}
\end{subfigure}
\caption{The results of our parameter scan in the $\mneu1$--$\tb$ plane
in the higgsino DM scenario for current (left) and anticipated future
limits (right) from \gmin2.
The color coding is as in \protect\reffi{mn1-mc1-higgsino}.
The black line indicates the current exclusion bounds for heavy MSSM
Higgs bosons at the LHC (see text). 
}
\label{mneu1-tb-higgsino}
\end{figure}

We finish our analysis of the higgsino DM case with the
$\mneu1$--$\tb$ plane presented in \reffi{mneu1-tb-higgsino} with the same
color coding as in \reffi{mn1-mc1-higgsino}. The \gmin2\ constraint is
fulfilled in a triangular region with the largest neutralino masses allowed
for the largest $\tb$ values (where we stopped our scan at $\tb = 60$),
following the analytic dependence of the \gmin2\ contributions
in \refse{sec:constraints}, $\amu \propto \tb/m_{\rm EW}^2$ (where we denote
with $m_{\rm EW}$ an overall EW mass scale.
In agreement with the previous plots, the largest values for the
lightest neutralino masses are $\sim 650 \gev$ $(\sim 600 \gev)$ for the
current (anticipated future) \gmin2\ constraint. The DM relic density
does not give any additional constraint. 
The points allowed by the DM DD limits (cyan and red) yield the 
observed reduction to about $\sim 500 \gev$. 
The LHC constraints cut out all points at low $\mneu1$, but nearly
independent of $\tb$. As observed before, they yield a small further
reduction in the case of the anticipated future \gmin2\ accuracy. 

In \reffi{mneu1-tb-higgsino} we also show as black lines the current bound
from LHC searches for heavy neutral Higgs
bosons~\cite{Bahl:2018zmf} in the channel $pp \to H/A \to \tau\tau$
in the $M_h^{125}(\tilde\chi)$ benchmark scenario
(based on the search data published
in \citere{Aad:2020zxo} using $139\, \ifb$.)%
\footnote{We thank T.~Stefaniak for the evaluation of this limit, using
the latest version of
{\tt HiggsBounds}~\cite{Bechtle:2008jh,Bechtle:2011sb,Bechtle:2013wla,Bechtle:2015pma,Bechtle:2020pkv}.}%
.
~In this scenario light
charginos and neutralinos are present, suppressing the $\tau\tau$ decay
mode and thus yielding relatively weak limits in the $\MA$--$\tb$ plane
(see, e.g., Fig.~5 in~\cite{Bahl:2018zmf}). The black lines correspond to
$\mneu1 = \MA/2$, i.e.\ roughly to the requirement for $A$-pole
annihilation, where points
above the black lines are experimentally excluded. 
It can be observed that all points allowed by \gmin2\ are above the
exclusion curve. This renders the effects of $A$-pole annihilation in
this scenario effectively irrelevant (and justifies our choice to
fix $\MA$ above the TeV scale).


\subsection{Wino DM}
\label{sec:wino}

The next case under investigation is the wino DM case, as discussed
in \refse{sec:scan}. We follow the analysis flow as described
in \refse{sec:flow} and denote the points surviving certain constraints
with different colors as defined in \refse{sec:higgsino}.
The vacuum stability test had no effect on the points passing all
other constraints.

\begin{figure}[htb!]
\centering
\begin{subfigure}[b]{0.48\linewidth}
\centering\includegraphics[width=\textwidth]{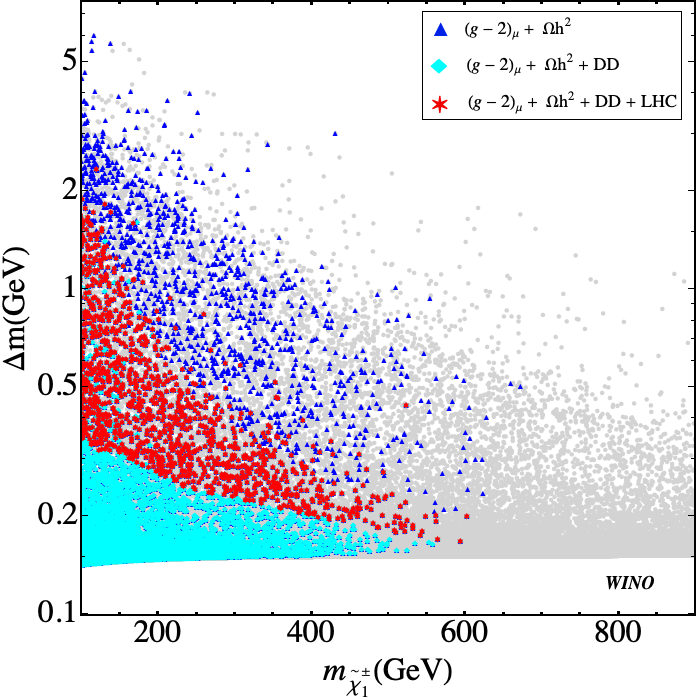}
\caption{}
\label{}
\end{subfigure}
~
\begin{subfigure}[b]{0.48\linewidth}
\centering\includegraphics[width=\textwidth]{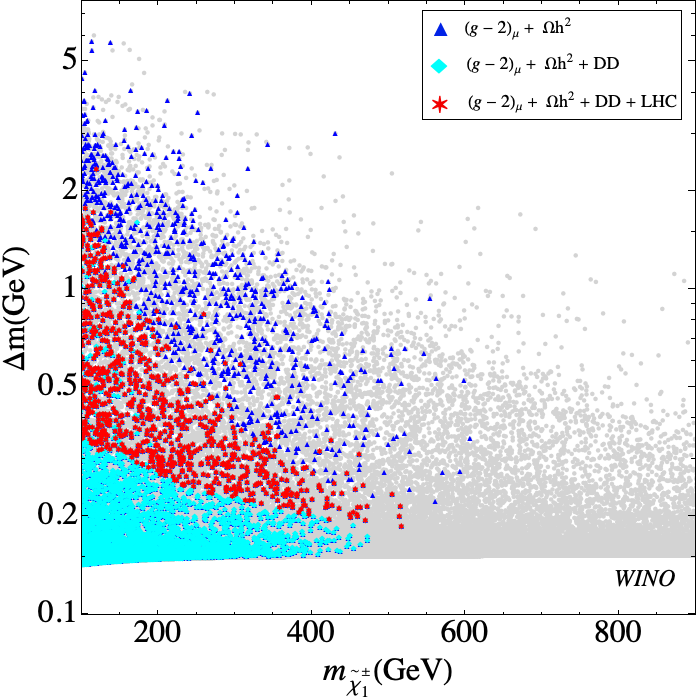}
\caption{}
\label{}
\end{subfigure}
\caption{The results of our parameter scan in the $\mcha1-\Delta m (=
m_{\cha1}-m_{\neu1})$ plane 
for the wino DM scenario for current (left) and anticipated future
limits (right) from \gmin2.
The color coding is as in \protect\reffi{mn1-mc1-higgsino}. 
}
\label{mc1-delm-wino}
\end{figure}

\begin{figure}[htb!]
	\vspace{2em}
\centering
\begin{subfigure}[b]{0.48\linewidth}
\centering
\includegraphics[width=\textwidth]{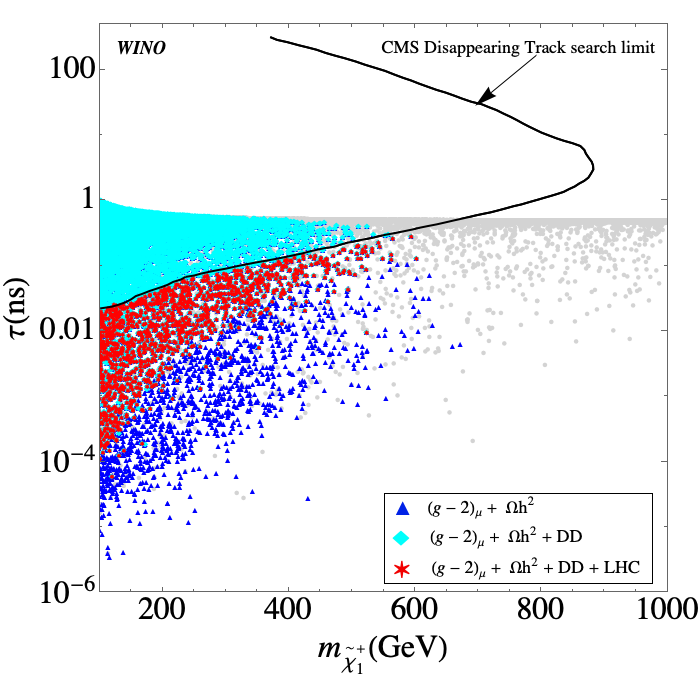}
	\caption{}
	\label{}
\end{subfigure}
~
\begin{subfigure}[b]{0.48\linewidth}
\centering\includegraphics[width=\textwidth]{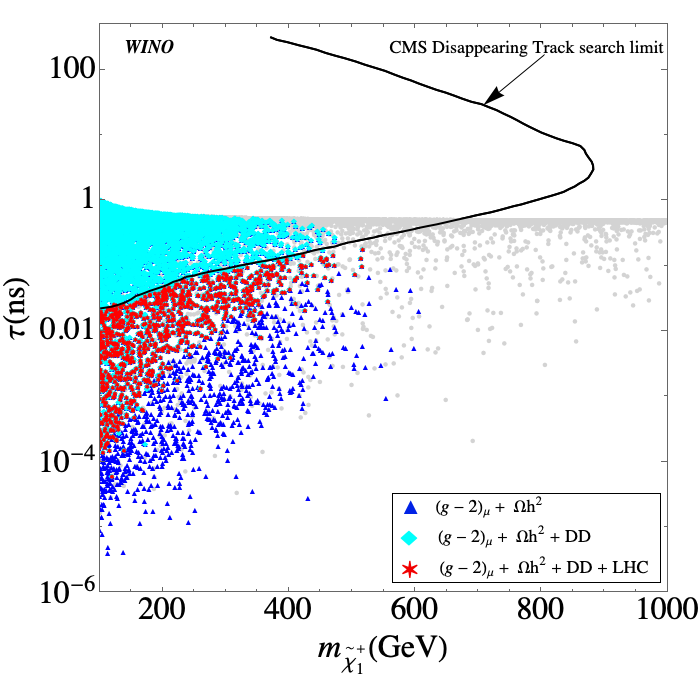}
	\caption{}
	\label{}
\end{subfigure}
\caption{The results of our wino parameter scan in the $\mcha1$--$\tau$
plane for current (left) and anticipated future
limits (right) from \gmin2, where $\tau$ is the lifetime
of the chargino decaying to $\pi^\pm\neu1$.
The current limit from CMS~\cite{Sirunyan:2020pjd} is shown as the
black solid line. 
The color coding is as in \protect\reffi{mc1-delm-wino}.}
\label{mc1-tau-wino}
\end{figure}

In \reffi{mc1-delm-wino} we show our results in the
$\mcha1$--$\De m (= \mcha1 - \mneu1)$
plane for the current (left) and future (right) \gmin2\ constraint,
see \refeqs{gmt-diff} and (\ref{gmt-fut}), respectively.
We display the results for $\De m$ rather than for $\mcha1$, since the
mass difference is very small, and the various features are more
easily visible in this plane. It should be remembered that we have applied the
one-loop shift to, in particular, $\mneu1$ that allows for the decay
$\cha1 \to \neu1 \pi^\pm$, see the discussion
in \refse{sec:scan}. This results in the lower bound on $\De m$ of
$\sim 0.15 \gev$. 
As in the higgsino scenario, see \refse{sec:higgsino}, all points that
pass the \gmin2\ constraint (current and anticipated future) also pass
the relic density constraint, shown as blue triangles
in \reffi{mc1-delm-wino}. The highest allowed chargino masses are
bounded from above by the \gmin2\ constraint. 
The overall allowed parameter space, shown as red stars, is
furthermore bounded
from ``above'' by the  DD limits and from ``below'' by the LHC
constraints. The DD limits cut away larger mass differences,
which can be understood as follows. 
The $h\neu1\neu1$ coupling for a wino-like $\neu1$
is given by~\cite{Hisano:2004pv}
\begin{eqnarray}
c_{h\neu1\neu1}\simeq
\frac{\MW}{M_2^2-\mu^2}(M_2+\mu\sin2\beta),
\label{ddwino}
\end{eqnarray}
in the limit of $||\mu|-M_2|\gg \MZ$ and a decoupled $\cp$-odd Higgs
boson (assuming also that the $h$-exchange dominates over the
$H$~contribution in the (spin independent) DD bounds). 
This coupling becomes large for $\mu \sim M_2$. On the other hand,
the tree level mass splitting between the wino-like states
$\chapm1$ and $\neu1$ generated (mainly by the mixing of the lighter
chargino with the charged higgsino) is given as~\cite{Ibe:2012sx}
\begin{eqnarray}
\De m (= \mcha1 - \mneu1) \simeq
\frac{ \MW^4 (\sin 2\beta)^2\tan^2 \theta_{\mathrm{w}} }{ (M_1 - M_2) \mu^2 },
\label{delmtree}
\end{eqnarray}
for $|M_1 - M_2| \gg \MZ$. The masss splitting increases for
smaller $\mu$ values and thus coincides with larger DD cross sections,
as discussed with \refeq{ddwino}.
All limits together yield maximum $\De m \sim 2 (0.2) \gev$ for
$\mcha1 \sim 100 (600) \gev$ for the current \gmin2\ constraint. The
upper limit is reduced to $\sim 500 \gev$ for the future anticipated 
\gmin2\ constraint.

The relevant LHC constraint is further analyzed
in \reffi{mc1-tau-wino}, where we show the plane $\mcha1$--$\tau_{\cha1}$.
$\tau_{\cha1}$ denotes the lifetime of the chargino decaying to
$\pi^\pm \neu1$.
Overlaid as black line is the bound from (CMS) charged disappearing track
analysis~\cite{Sirunyan:2020pjd}. One can observe that this constraint,
cutting out parameter points between $\tau_{\cha1} \sim 0.01$~ns and
$\sim 1$~ns is responsible for the main LHC exclusion.
It should be noted that in the ``red star area'' also some points
appear cyan, i.e.\ excluded by (other) LHC searches, where the most
relevant channels are pair production of sleptons leading
to two leptons and $\met$ in the final state~\cite{Aad:2019vnb}.
However, these channels are not strong enough to exclude more of
the $\mcha1$--$\tau_{\cha1}$ plane than the disappearing track
search. It can be expected in the (near) future that improved DD bounds,
cutting the allowed parameter space from small $\cha1$ lifetime and
improved disappearing track searches, cutting from large lifetimes,
may substantially shrink the allowed parameter space, sharpening the
upper limit on $\mcha1$ and the prospects for future collider
searches. The two bounds together have the potential to firmly rule
out the case of wino DM in the MSSM, will be discussed below.

\begin{figure}[htb!]
\vspace{2em}
\centering
\begin{subfigure}[b]{0.48\linewidth}
\centering\includegraphics[width=\textwidth]{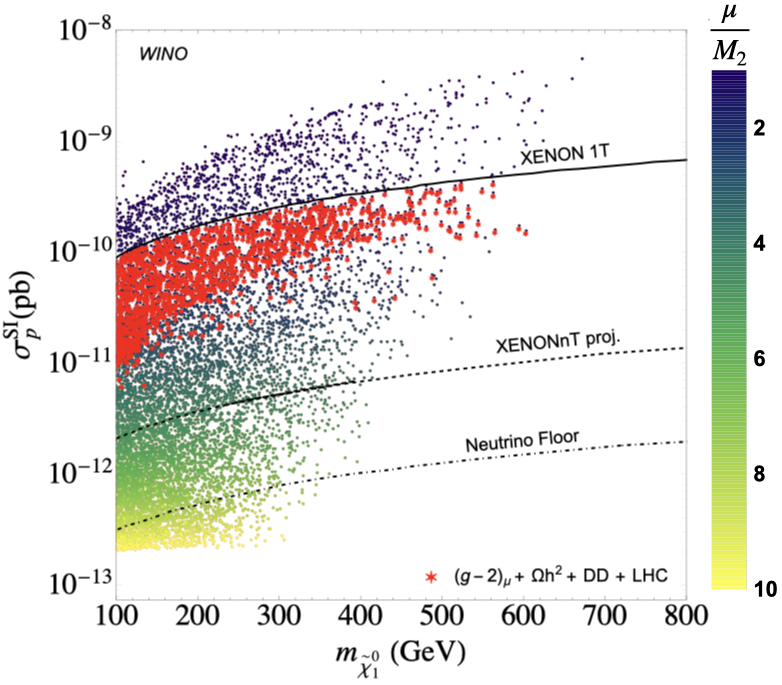}
	\caption{}
	\label{}
\end{subfigure}
~
\begin{subfigure}[b]{0.48\linewidth}
\centering\includegraphics[width=\textwidth]{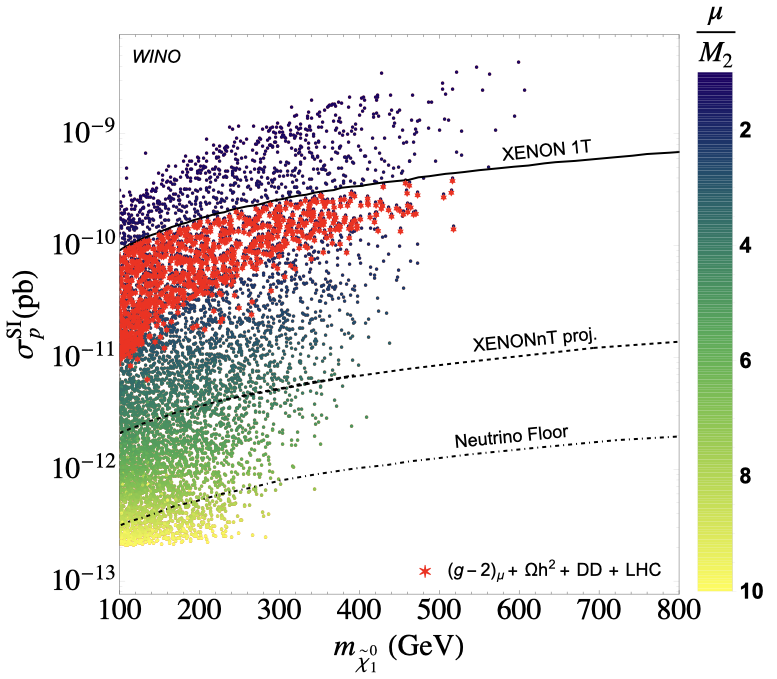}
	\caption{}
	\label{}
\end{subfigure}
\caption{Scan results in the $\mneu1$-$\ssi$ plane for the wino DM
scenario for current (left)
and anticipated future limits (right) from \gmin2. The color coding of
the points denotes $\mu/M_2$ and the black solid line indicates the current DD
limit from XENON-1T while the black dashed and dot-dashed lines are respectively
the projected reach of XENON-nT and coherent neutrino scattering
floor. 
In red we show the points fulfilling \gmin2\ , relic 
density, DD and additionally the LHC constraints.
}
\label{mneu1-ssi-wino}
\end{figure}

The impact of the DD experiments is demonstrated 
in \reffi{mneu1-ssi-wino}. We show the $\mneu1$--$\ssi$ plane for
current (left) and anticipated future limits (right) from \gmin2. 
The color coding of the points (from yellow to dark green) denotes $\mu/M_2$,
whereas in red we show the points fulfilling \gmin2\ , relic
density, DD and the LHC constraints.
The solid black line indicates the current DD limits, here
taken for sake of simplicity from XENON1T~\cite{XENON}, as discussed
in \refse{sec:constraints}. It can be seen that a slight downward shift of
this limit, e.g.\ due to additional DD experimental limits from
LUX~\cite{LUX} or PANDAX~\cite{PANDAX}, would not
change our results in a relevant way. However, moderately
improved limits may have a strong impact, as discussed above.
The scanned parameter space extends from large $\ssi$
values, given for the smallest scanned $\mu/M_2$ values to the
smallest ones, reached for the largest scanned $\mu/M_2$, i.e.\
the $\ssi$ constraints are particularly strong for small $\mu/M_2$. 
Given both CDM constraints and the LHC
constraints, shown in red, the smallest $\mu/M_2$ value we find
is~1.5 for both the current and
anticipated future \gmin2\ bound.
As mentioned above, the DD bound can become relevantly stronger
with future experiments. We show
as dashed line the projected limit of
XENONnT~\cite{Aprile:2020vtw}, and the dot-dashed line indicates the
neutrino floor~\cite{Ruppin:2014bra}.
One can see that the XENONnT result will
either firmly exlude or detect a wino DM candidate, possibly in
conjunction with improved disappearing track searches at the LHC, as
discussed above.

The distribution of the lighter slepton mass (where it should be kept in
mind that we have chosen the same masses for all three generations,
see \refse{sec:model}) is presented in the $\mneu1$--$\msl1$ plane
in \reffi{mneu1-mslep-wino}, with the same color coding as
in \reffi{mc1-delm-wino}.
The \gmin2\ constraint places important constraints in this mass plane,
since both types of masses enter into the contributing SUSY diagrams,
see \refse{sec:constraints} (and obviously all points pass the DM
relic density upper limit). 
The \gmin2\ constraint is satisfied in a triangular region with its tip
around $(\mneu1, \msl1) \sim (700 \gev, 700 \gev)$ in the case of
current \gmin2\ constraints, and around $\sim (600 \gev, 700 \gev)$ in
the case of the anticipated future limits. The highest slepton masses
reached are about $1500 (1200) \gev$, respectively; i.e.\ the impact of the
anticipated improved limits is clearly visible as an {\it upper} limit. 
After including the DD and LHC constraints the upper limits for the
LSP are slightly reduced by $\sim 100 \gev$, whereas the upper limits on
sleptons are not affected. Concerning the LHC searches, the
constraints from slepton
pair production searches rule out some parameter points from the low $\msl1$
region. Larger slepton masses are excluded by the same search if
the ``second'' slepton turns out to be relatively light. Points excluded
by compressed spectra searches~\cite{Aad:2019qnd}, depending on the
lightest chargino mass, can be found all over the plane.

\begin{figure}[htb!]
\centering
\begin{subfigure}[b]{0.48\linewidth}
\centering\includegraphics[width=\textwidth]{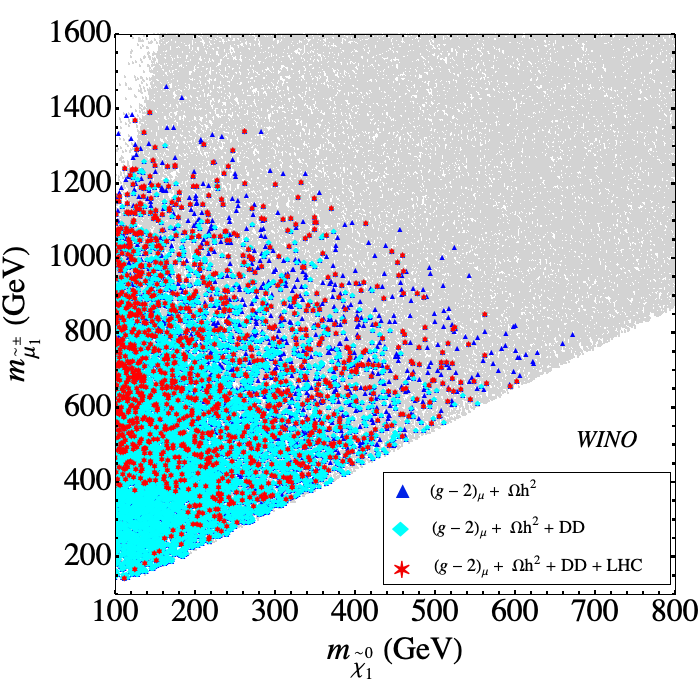}
\caption{}
\label{}
\end{subfigure}
~
\begin{subfigure}[b]{0.48\linewidth}
\centering\includegraphics[width=\textwidth]{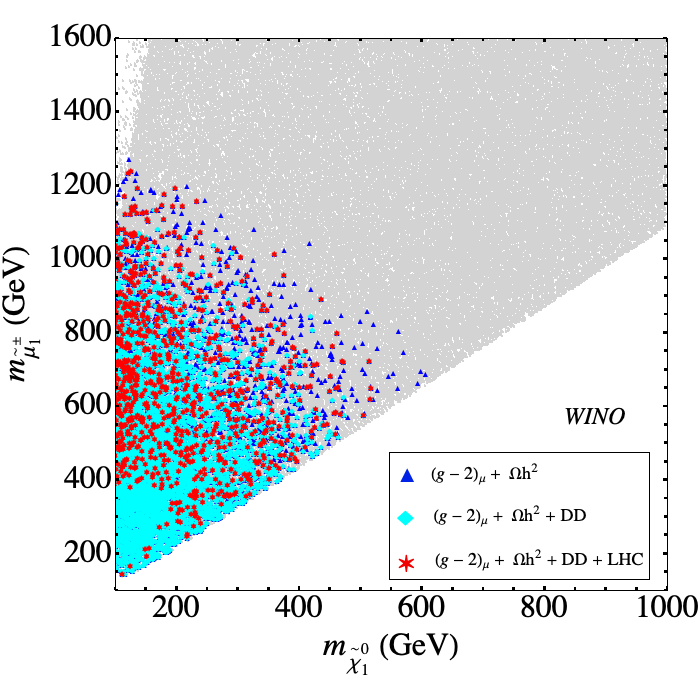}
\caption{}
\label{}
\end{subfigure}
\caption{The results of our parameter scan in the
	$\mneu1$--$\msmu1$ plane
	for the wino DM scenario for current (left) and
	anticipated future limits (right) from \gmin2.
	The color coding is as in \protect\reffi{mc1-delm-wino}.
}
\label{mneu1-mslep-wino}
\end{figure}

\begin{figure}[htb!]
\vspace{3em}
\centering
\begin{subfigure}[b]{0.48\linewidth}
\centering\includegraphics[width=\textwidth]{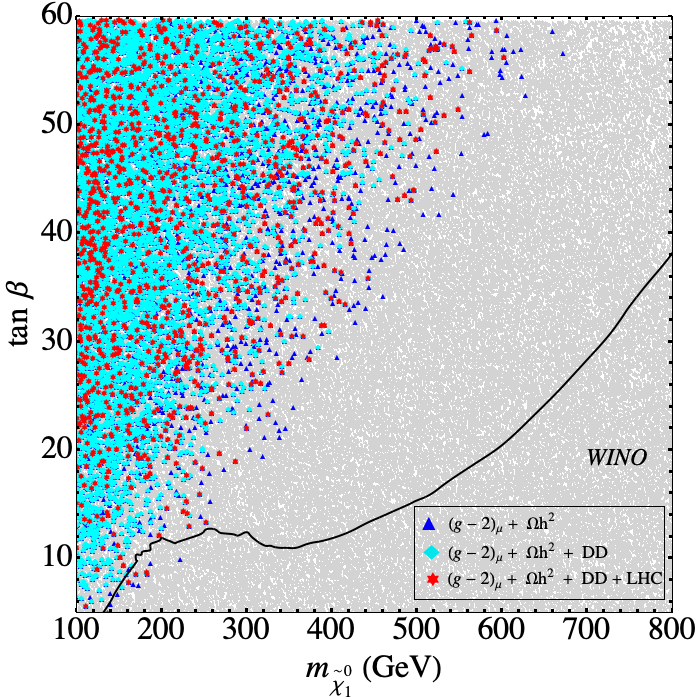}
\caption{}
\label{}
\end{subfigure}
~
\begin{subfigure}[b]{0.48\linewidth}
\centering\includegraphics[width=\textwidth]{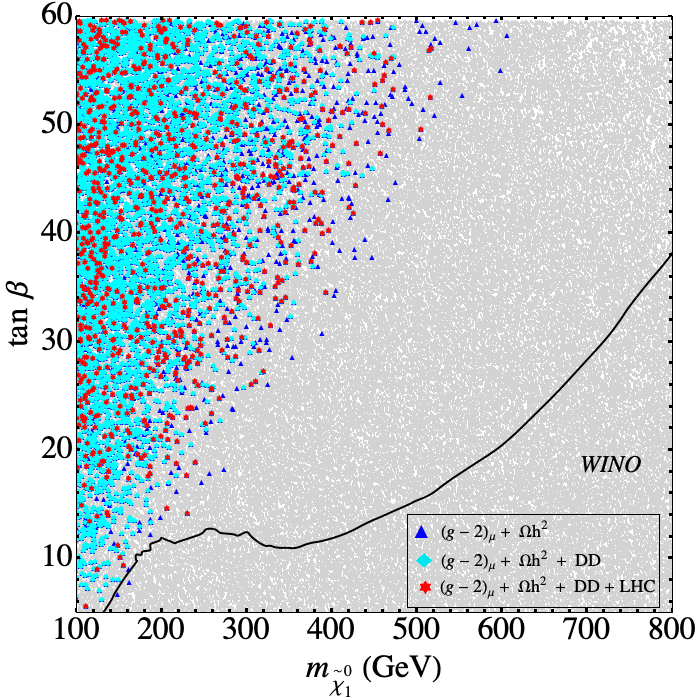}
\caption{}
\label{}
\end{subfigure}
\caption{The results of our parameter scan in the $\mneu1$-$\tb$ plane
in the wino DM scenario for current (left) and anticipated future
limits (right) from \gmin2.
The color coding is as in \protect\reffi{mc1-delm-wino}.
The black line indicates the current exclusion bounds for heavy MSSM
Higgs bosons at the LHC (see text).
}
\label{mneu1-tb-wino}
\end{figure}

We finish our analysis of the wino DM case with the
$\mneu1$-$\tb$ plane presented in \reffi{mneu1-tb-wino} with the same
color coding as in \reffi{mc1-delm-wino}. The \gmin2\ constraint is
fulfilled in a triangular region with largest neutralino masses allowed
for the largest $\tb$ values (where we stopped our scan at $\tb = 60$),
following the analytic dependence of the \gmin2\ contributions
in \refse{sec:constraints}, $\amu \propto \tb/m_{\rm EW}^2$ (where we denote
with $m_{\rm EW}$ an overall EW mass scale.
In agreement with the previous plots, the largest values for the
lightest neutralino masses are $\sim 600 \gev$ $(\sim 500 \gev)$ for the
current (anticipated future) \gmin2\ constraint. 
The points allowed by the DM constraints (blue/cyan) are distributed
all over the allowed region. 
The LHC constraints also cut out points distributed all over the
allowed triangle, in agreement with the previous discussion.

In \reffi{mneu1-tb-wino} we also show as black lines the current bound
from LHC searches for heavy neutral Higgs
bosons~\cite{Aad:2020zxo,Bahl:2018zmf}, see the discussion
in \refse{sec:higgsino}. Points
above the black lines are experimentally excluded. 
There are a few points passing the current \gmin2\ constraint
below the black $A$-pole line, reaching up to $\mneu1 \sim 220 \gev$,
for which the $A$-pole annihilation could provide the correct DM relic
density. For the anticipated future accuracy in \gmin2\ this mechanism
would effectively be absent, making the $A$-pole annihilation in this
scenario marginal.


\subsection{Bino/wino and bino DM}
\label{sec:bino}

In this section we analyze the case of bino/wino and bino DM, as defined
in \refse{sec:scan}. The three cases defined there correspond exactly to
the set of analyses in \citere{CHSold}. However, we now apply the DM
relic density as an upper bound (``DM upper bound''), whereas
in \citere{CHSold} the LSP was required to give the full amount of CDM
(``DM full''). Since overall the results are similar to the ones
found in \citere{CHSold}, we will keep the discussion brief, but try
to highlight the differences w.r.t.\ \citere{CHSold}.
For all scenarios we find that the vacuum
stability bounds have no impact on the final mass limits found.

\subsubsection{Bino/wino DM with \boldmath{$\cha1$-coannihilation}}
\label{sec:chaco}

In \reffi{mn1-mc1-chaco} we show our results in the $\mneu1$--$\mcha1$
plane for the current (left) and future (right) \gmin2\ constraint,
see \refeqs{gmt-diff} and (\ref{gmt-fut}), respectively.
The color coding is defined in \refse{sec:higgsino}. By definition
of $\cha1$-coannihilation the points are clustered in the diagonal of
the plane. Overall we observe here exactly the same pattern of points as in
the case of ``DM full''~\cite{CHSold}. After taking into account all
constraints we find upper limits of $\sim 600 (500) \gev$ in the case of
the current (future) \gmin2\ limits.
Thus, the experimental data set about the same upper as well as lower
bounds as in \citere{CHSold} (modulo differences due to point density 
artefacts). Consequently, the search targets for the upcoming LHC runs 
particular for future $e^+e^-$ colliders remain about the same.

\begin{figure}[htb!]
\centering
\begin{subfigure}[b]{0.48\linewidth}
\centering\includegraphics[width=\textwidth]{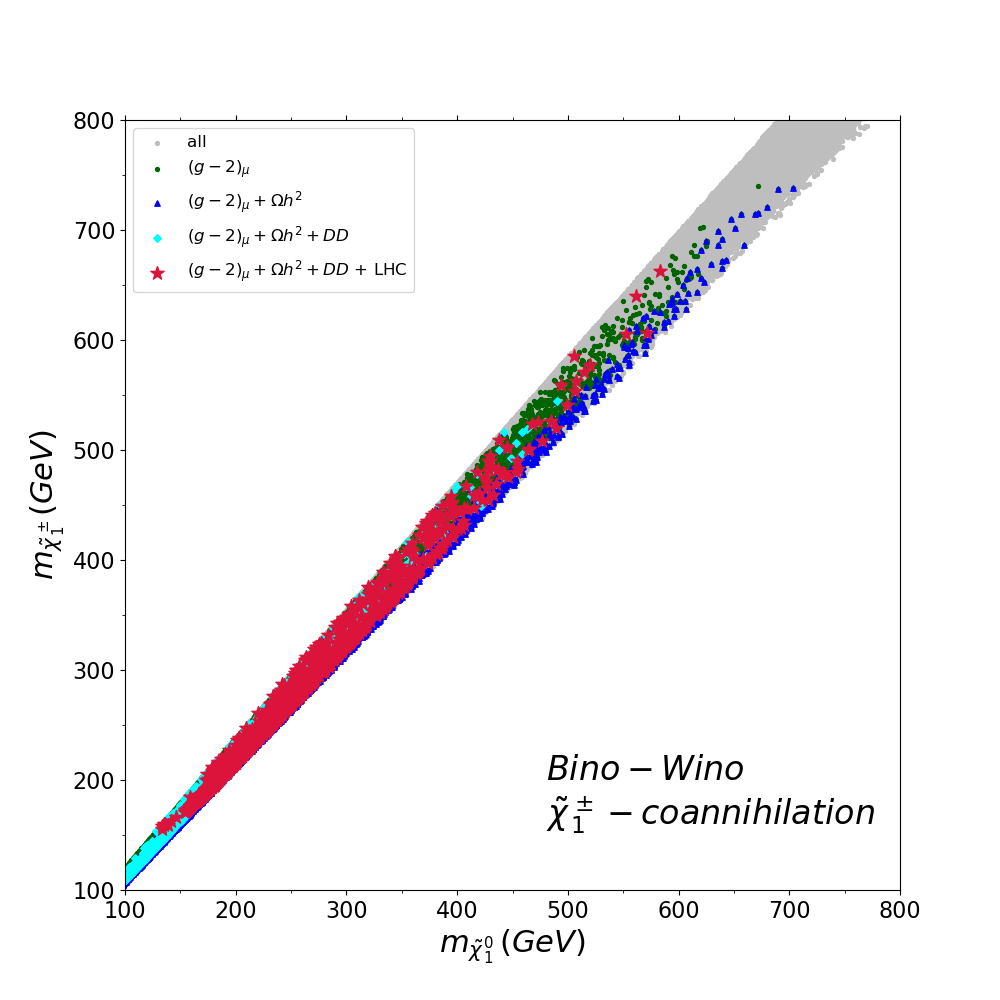}
\caption{}
\label{}
\end{subfigure}
~
\begin{subfigure}[b]{0.48\linewidth}
\centering\includegraphics[width=\textwidth]{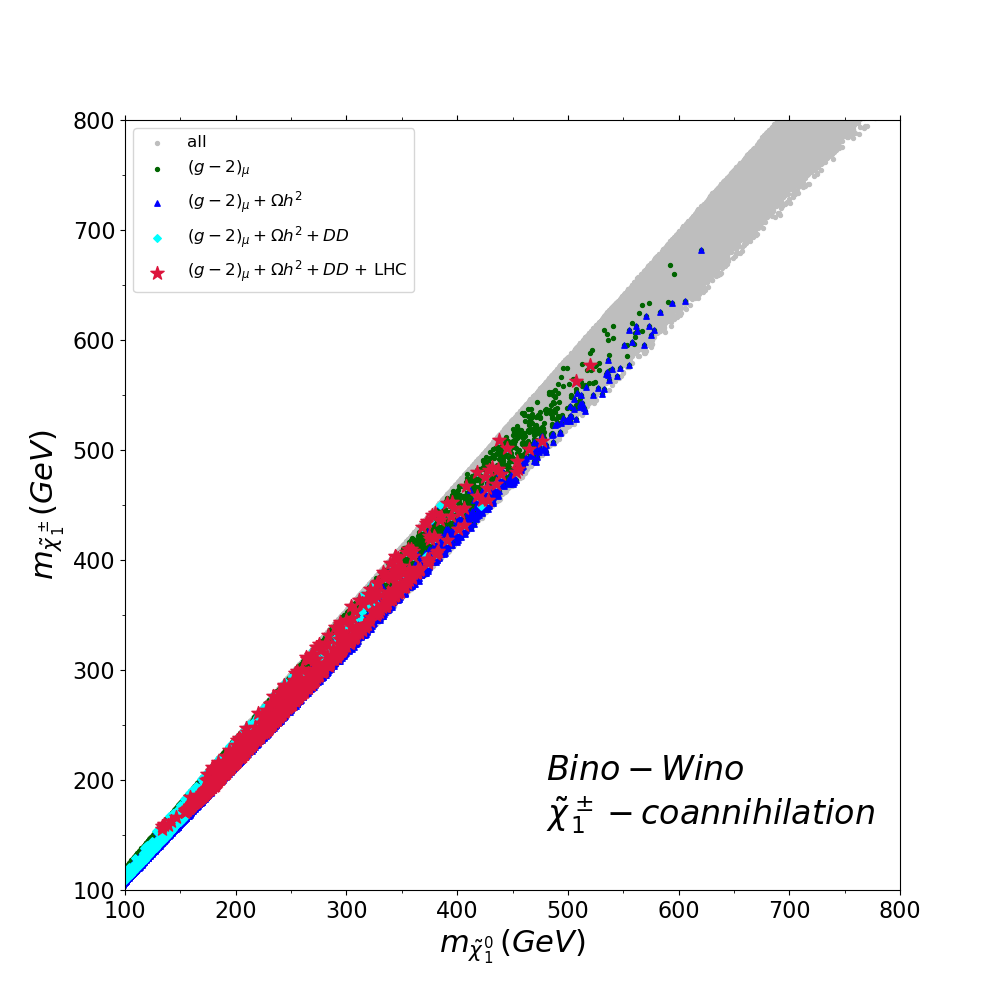}
\caption{}
\label{}
\end{subfigure}
\caption{The results of our parameter scan in the $\mneu1-\mcha1$ plane
for the bino-wino $\cha1$-coannihilation scenario for current (left) and
anticipated future 
limits (right) from \gmin2. For the color coding: see text.
}
\label{mn1-mc1-chaco}
\end{figure}

The impact of the DD experiments is demonstrated 
in \reffi{mneu1-ssi-chaco}. We show the $\mneu1$--$\ssi$ plane for
current (left) and anticipated future limits (right) from \gmin2. 
The color coding of the points (from yellow to dark green) denotes $\mu/M_1$,
whereas in red we show the points fulfilling all constraints including
the LHC ones. As in \reffi{mneu1-ssi-higgsino} the solid black line indicates
the current DD limits from XENON1T~\cite{XENON}.
Also here the results are in very good agreement with \citere{CHSold}
(again modulo point density artefacts).
The scanned parameter space extends from large $\ssi$
values, given for the smallest scanned $\mu/M_1$ values to the
smallest ones, reached for the largest scanned $\mu/M_1$, i.e.\
the $\ssi$ constraints are particularly strong for small $\mu/M_1$. 
Given in red the points fulfilling \gmin2\ , both CDM constraints and the LHC
constraints, the smallest $\mu/M_1$ value we find
is~1.67 for the current and~1.78 for the
anticipated future \gmin2\ bound.
The dashed line indicates the projected XENONnT
limit~\cite{Aprile:2020vtw}, and the dot-dashed line indicates the 
neutrino floor~\cite{Ruppin:2014bra}. One can see that XENONnT will
not be able to fully test the chargino co-annihilation scenario, with
some points that pass all constraints (red) being even below the
neutrino floor.

\begin{figure}[htb!]
\vspace{2em}
\centering
\begin{subfigure}[b]{0.48\linewidth}
\centering\includegraphics[width=\textwidth]{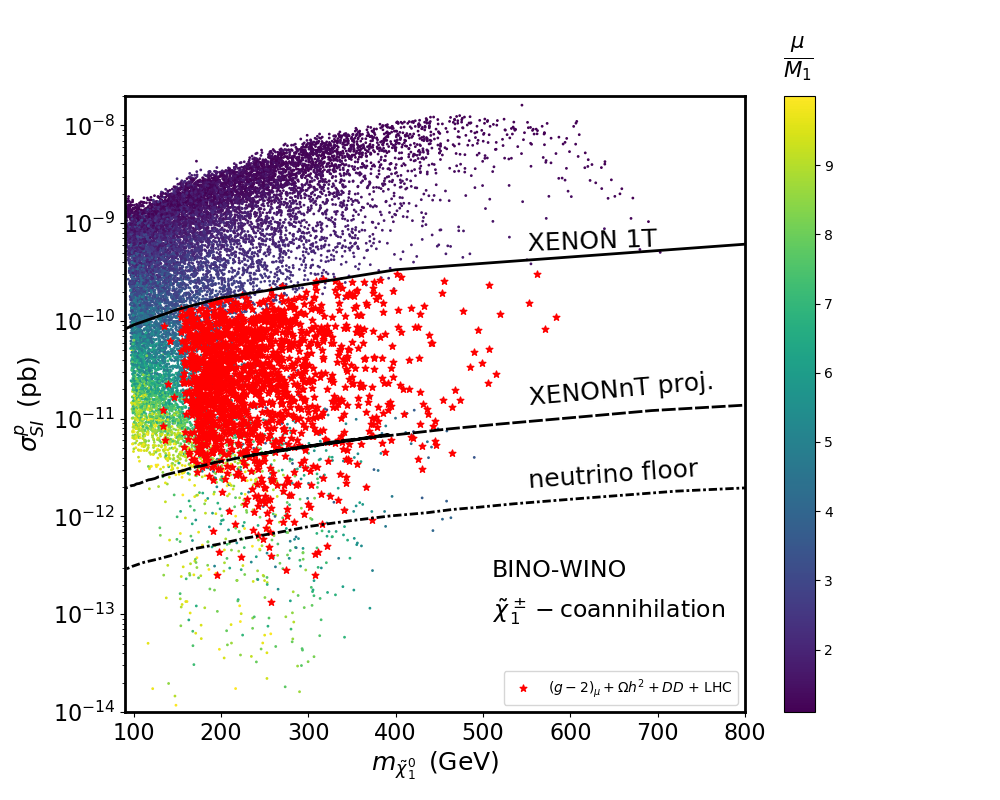}
\caption{}
\label{}
\end{subfigure}
~
\begin{subfigure}[b]{0.48\linewidth}
\centering\includegraphics[width=\textwidth]{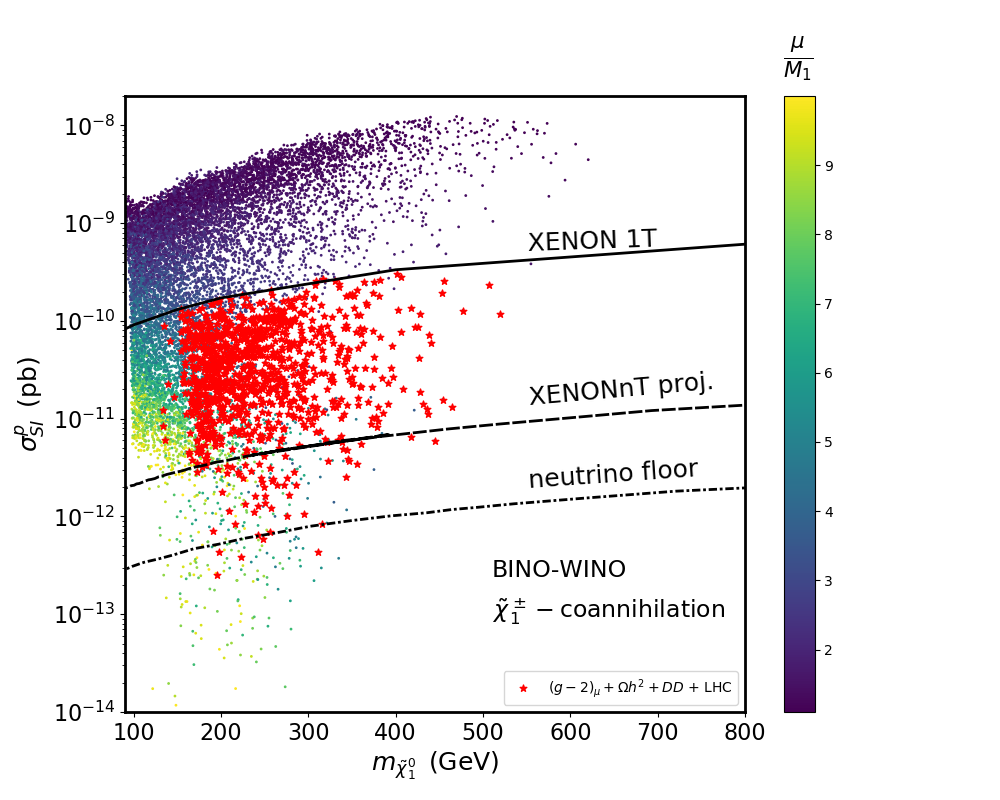}
\caption{}
\label{}
\end{subfigure}
\caption{Scan results in the $\mneu1$--$\ssi$ plane for bino-wino
$\cha1$-coannihilation scenario for current (left)
and anticipated future limits (right) from \gmin2. The color coding of
the points denotes $\mu/M_1$ and the black lines indicates the DD
limits (see text). In red we show the points fulfilling \gmin2\ , the relic
density, DD and the LHC constraints. 
}
\label{mneu1-ssi-chaco}
\end{figure}

The distribution of the lighter slepton mass (where it should be kept in
mind that we have chosen the same masses for all three generations,
see \refse{sec:model}) is presented in the $\mneu1$--$\msl1$ plane
in \reffi{mneu1-mslep-chaco}, with the same color coding as
in \reffi{mn1-mc1-chaco}.
The \gmin2\ constraint places important constraints in this mass plane,
since both types of masses enter into the contributing SUSY diagrams,
see \refse{sec:constraints}. 
The constraint is satisfied in a triangular region with its tip
around $(\mneu1, \msl1) \sim (700 \gev, 800 \gev)$ in the case of
current \gmin2\ constraints, and around $\sim (600 \gev, 700 \gev)$ in
the case of the anticipated future limits, i.e.\ the impact of the
anticipated improved limits is clearly visible as an {\it upper} limit.
These results remain unchanged w.r.t.\ the ``DM full''
case~\cite{CHSold}. 
The points fulfilling the DM relic density constraint
(blue/cyan/red) are distributed 
all over the \gmin2\ allowed range. They extend to somewhat higher
slepton masses as compared to \citere{CHSold}, due to the less
restrictive DM upper bound.%
\footnote{
A very few points have at higher $\msl1$ have the ``correct'' relic
density, which can be attributed to a substantially large sample of points.}%
~The DD limits cut away the largest values, in particular for
$\msl1$, which can be understood as follows. Large $\msl1$ values, with
correspondingly large sneutrino masses, require smaller $\mu$ to satisfy
the \gmin2\ constraint. This in turn puts them in tension with the DD
bounds, see \reffi{mneu1-ssi-chaco}.

\begin{figure}[htb!]
	\vspace{1em}
\centering
\begin{subfigure}[b]{0.48\linewidth}
\centering\includegraphics[width=\textwidth]{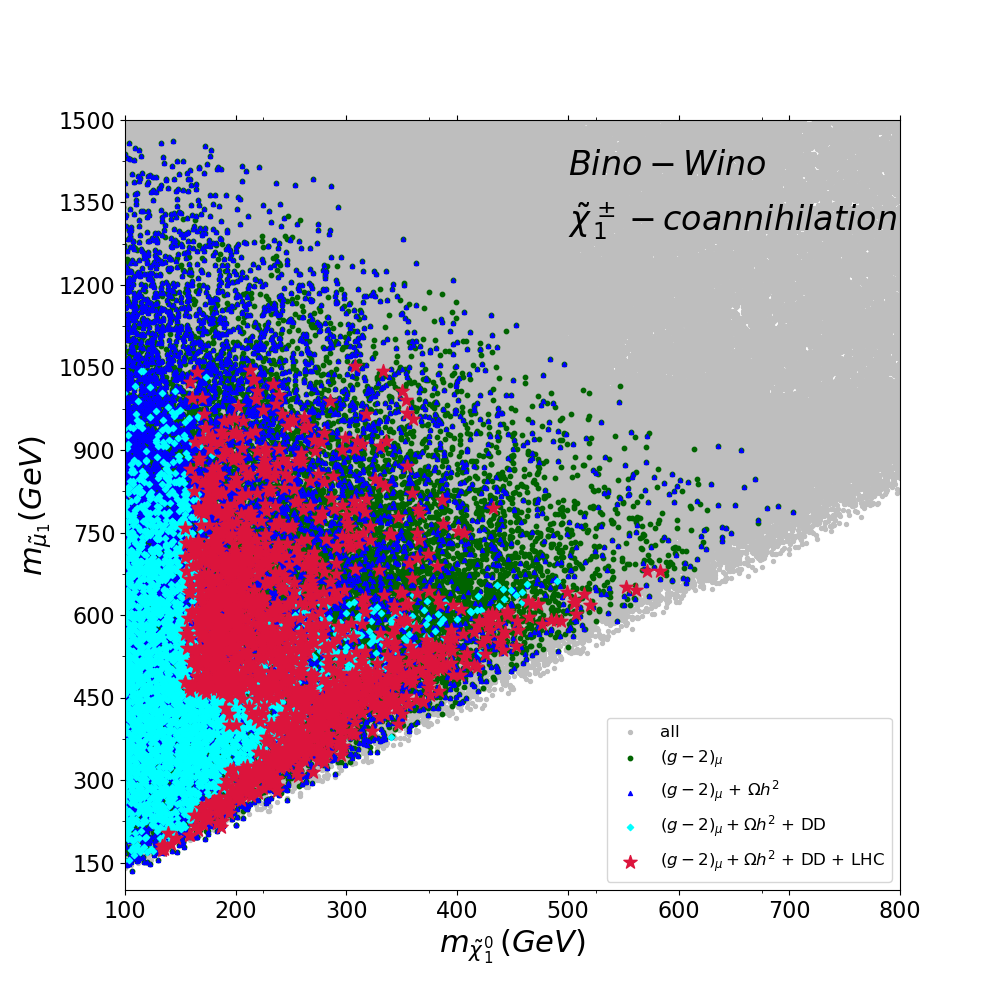}
	\caption{}
	\label{}
\end{subfigure}
~
\begin{subfigure}[b]{0.48\linewidth}
\centering\includegraphics[width=\textwidth]{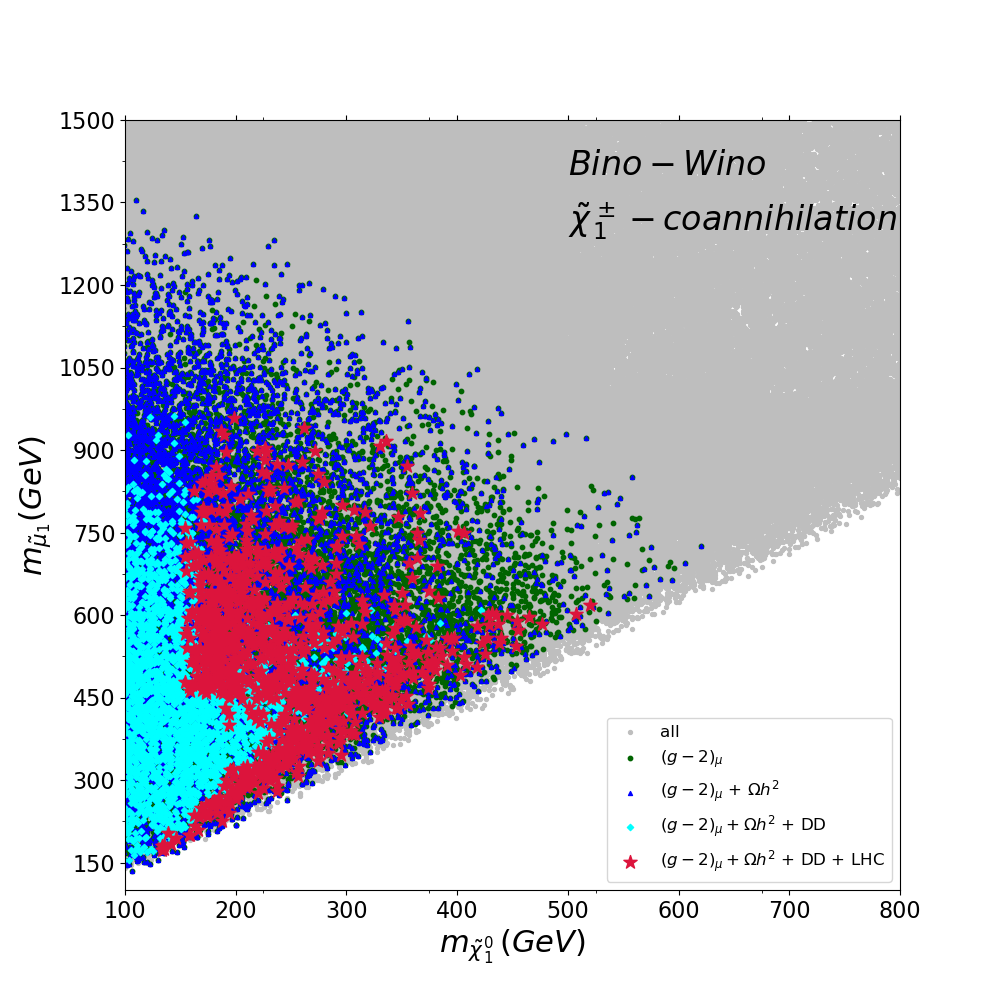}
	\caption{}
	\label{}
\end{subfigure}
\caption{The results of our parameter scan in the
$\mneu1$--$\msl1$ plane
for the bino-wino $\cha1$-coannihilation scenario for current (left) and
anticipated future limits (right) from \gmin2.
The color coding is as in \protect\reffi{mn1-mc1-chaco}.
}
\label{mneu1-mslep-chaco}
\end{figure}

The LHC constraints cut out lower slepton masses, following the same
pattern as in \citere{CHSold}. They cut away masses up to 
$\msl1 \lsim 450 \gev$, as well as part of the very low $\mneu1$
points nearly independent of $\msl1$. Here the latter ``cut'' is due to the
searches for compressed spectra with $\cha1, \neu2$ decaying
via off-shell gauge bosons~\cite{Aad:2019qnd}. 
The first ``cut'' is mostly a result of the searches for slepton
pair production with a decay to two leptons plus missing
energy~\cite{Aad:2019vnb}. 
As was demonstrated and discussed in detail in \citere{CHSold} for this
limit it is crucial to employ a proper re-cast of the LHC searches,
rather than a naive application of the published bounds,
Overall we can place an upper limit on the light slepton
mass of about $\sim 1050 \gev$ and $\sim 950\gev$ for the current
and the anticipated future accuracy of \gmin2, respectively.

\begin{figure}[htb!]
\centering
\begin{subfigure}[b]{0.48\linewidth}
\centering\includegraphics[width=\textwidth]{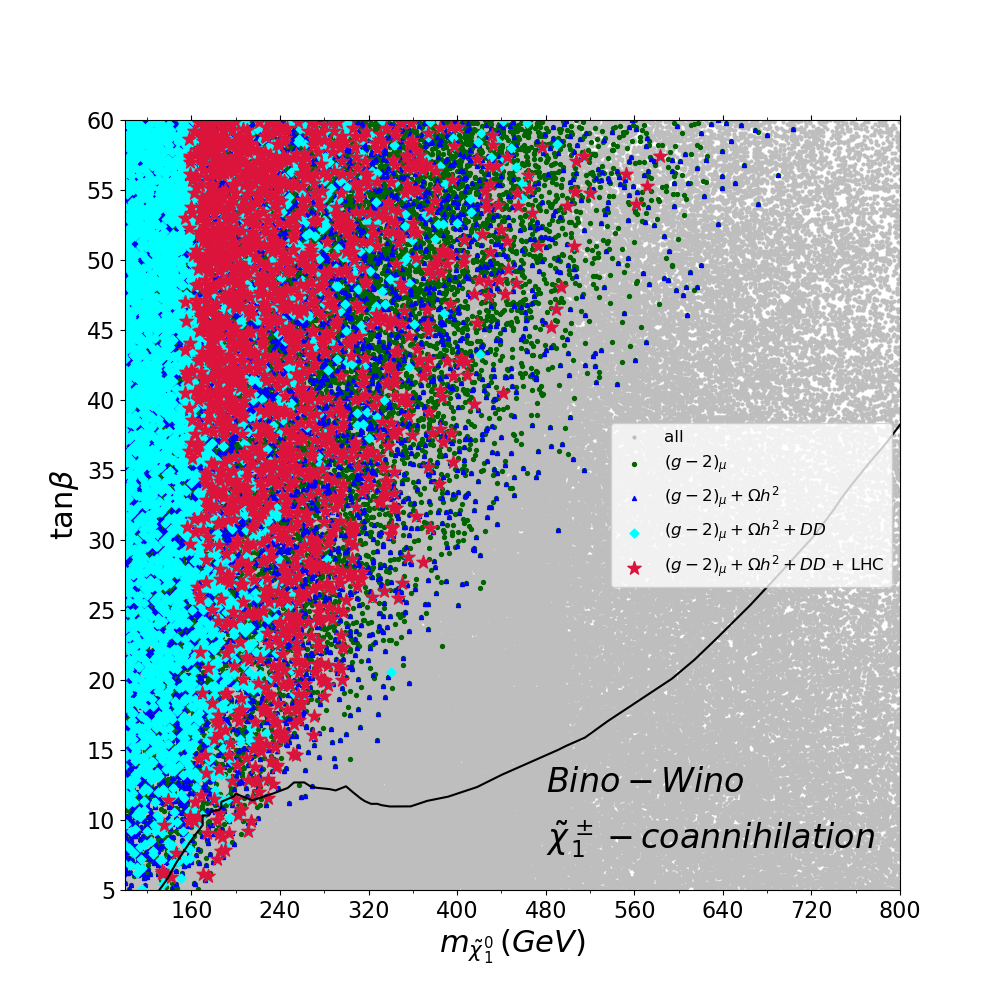}
	\caption{}
	\label{}
\end{subfigure}
~
\begin{subfigure}[b]{0.48\linewidth}
\centering\includegraphics[width=\textwidth]{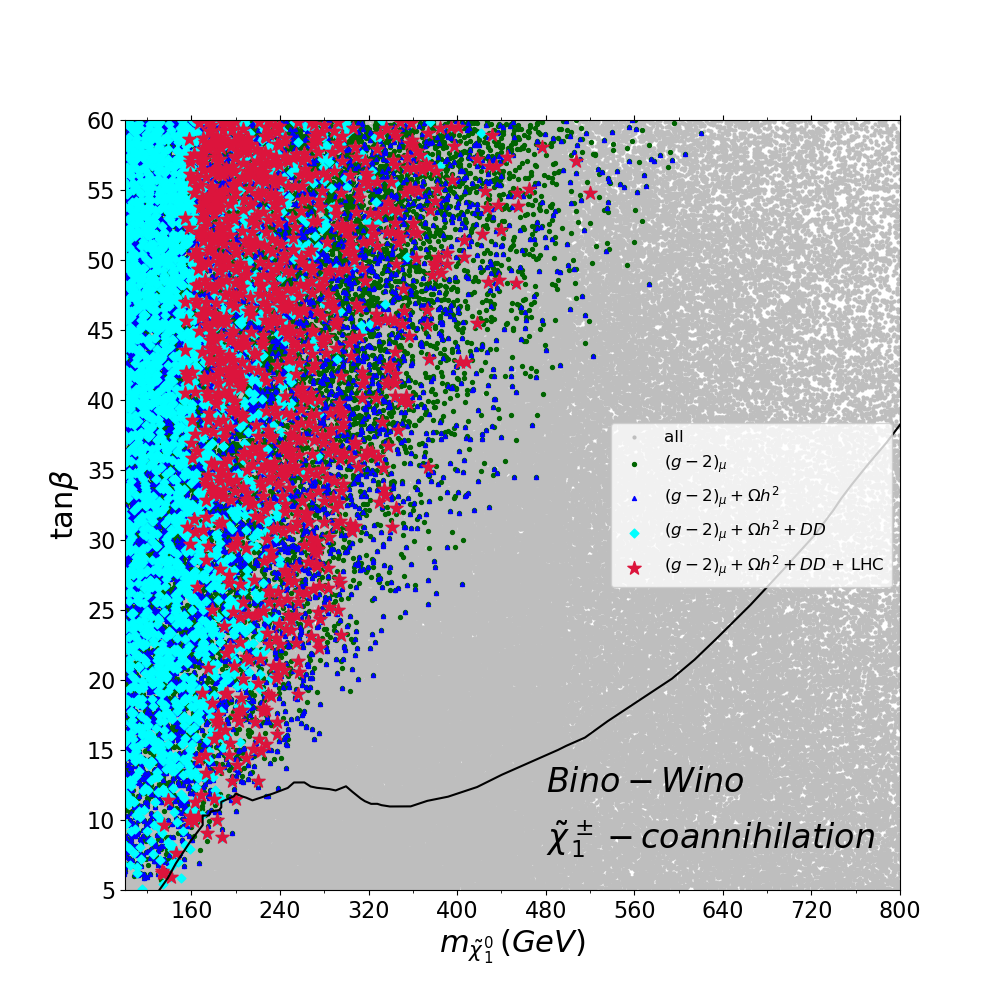}
	\caption{}
	\label{}
\end{subfigure}
\caption{The results of our parameter scan in the $\mneu1$--$\tb$ plane
in the bino-wino $\cha1$-coannihilation scenario for current (left) and
anticipated future limits (right) from \gmin2.
The color coding is as in \protect\reffi{mn1-mc1-chaco}.
The black line indicates the current exclusion bounds for heavy MSSM
Higgs bosons at the LHC (see text).}
\label{mneu1-tb-chaco}
\end{figure}

We finish our analysis of the $\cha1$-coannihilation case with the
$\mneu1$--$\tb$ plane presented in \reffi{mneu1-tb-chaco} with the same
color coding as in \reffi{mn1-mc1-chaco}.
For this plane we find that the results are in full agreement with the
``DM full'' case as analyzed in \citere{CHSold}. 
The \gmin2\ constraint is
fulfilled in a triangular region with largest neutralino masses allowed
for the largest $\tb$ values ($\tb = 60$).
In agreement with the previous plots, the largest values for the
lightest neutralino masses are $\sim 600 \gev$ $(\sim 500 \gev)$ for the
current (anticipated future) \gmin2\ constraint. 
The points allowed by the DM constraints (blue/cyan) are distributed
all over the allowed region. 
The LHC constraints cut out points at low $\mneu1$, but nearly
independent on $\tb$.

As in the previous scenarios, in \reffi{mneu1-tb-chaco} we also show as
black lines the current bound from LHC searches for heavy neutral Higgs
bosons~\cite{Bahl:2018zmf} in the channel $pp \to H/A \to \tau\tau$
in the $M_h^{125}(\tilde\chi)$ benchmark scenario.
As before the black lines correspond to
$\mneu1 = \MA/2$, i.e.\ roughly to the requirement for $A$-pole
annihilation, where points
above the black lines are experimentally excluded. The improved limits
of the experimental analysis based on
$139 \ifb$~\cite{Aad:2020zxo}, but now with the relaxed DM relic
density bound, still allow parameter points that cannot be regarded as
excluded. They are found for $\mneu1 \lsim 240 (200) \gev$ and
$\tb \lsim 12$ in the case
of the current (anticipated future) \gmin2\ constraints. To analyze these
points a dedicated analysis of the $A$-pole annihilation would be
required. This dedicated analysis we leave for future work.


\subsubsection{Bino DM with \boldmath{$\Slpm$}-coannihilation case-L}
\label{sec:slepL}

We now turn to the case of bino DM with $\Slpm$-coannihilation. As
discussed in in \refse{sec:scan} we distinguish two cases, depending
which of the two slepton soft SUSY-breaking parameters is set to be close to
$\mneu1$. We start with the Case-L, where we chose $\msl{L} \sim M_1$,
i.e.\ the left-handed 
charged sleptons as well as the sneutrinos are close in mass to the LSP.
We find that all six sleptons are close in mass and differ by less
than $\sim 50 \gev$.

\begin{figure}[htb!]
\centering
\begin{subfigure}[b]{0.48\linewidth}
\centering\includegraphics[width=\textwidth]{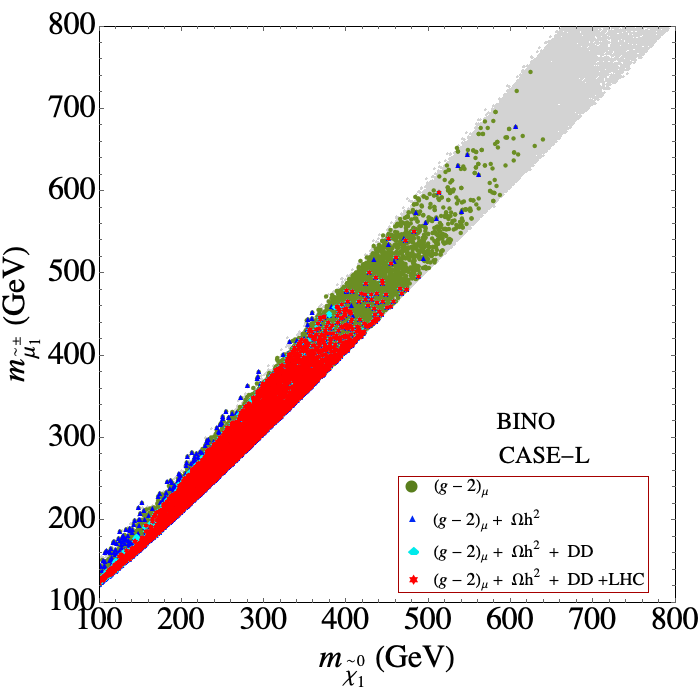}
\caption{}
\label{}
\end{subfigure}
~
\begin{subfigure}[b]{0.48\linewidth}
\centering\includegraphics[width=\textwidth]{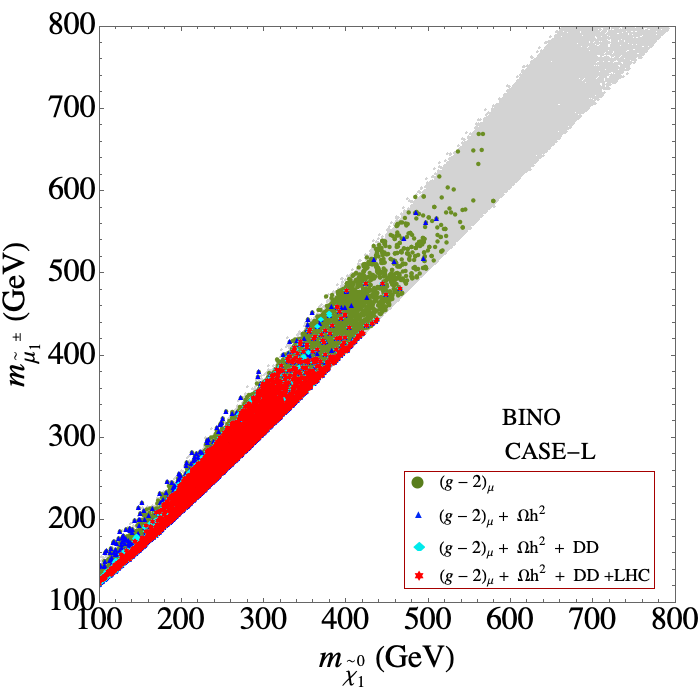}
\caption{}
\label{}
\end{subfigure}
\caption{The results of our parameter scan in the
$\mneu1$--$\msmu1$ plane
for the  $\Sl$-coannihilation case-L scenario for current (left) and
anticipated future limits (right) from \gmin2.
The color coding is as in \protect\reffi{mc1-delm-wino}.
}
\label{mneu1-msmu1-caseL}
\end{figure}

\begin{figure}[htb!]
\vspace{2em}
\centering
\begin{subfigure}[b]{0.48\linewidth}
\centering\includegraphics[width=\textwidth]{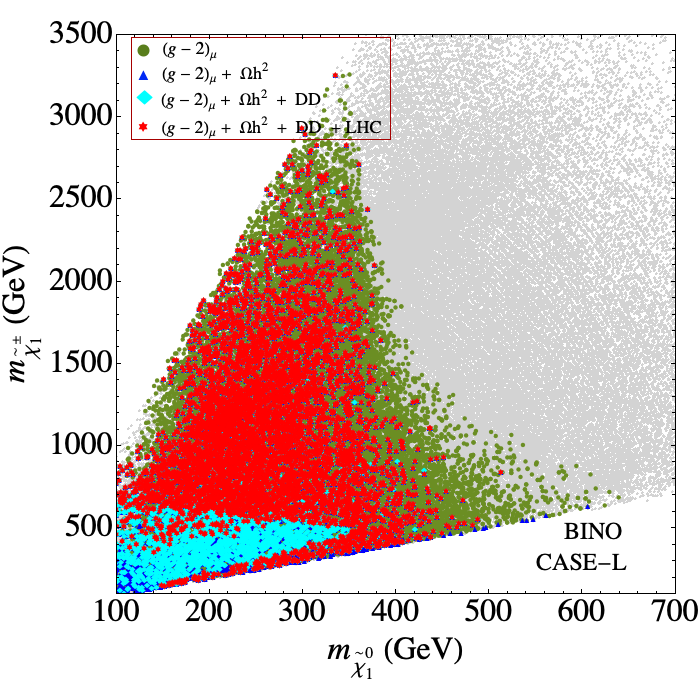}
\caption{}
\label{}
\end{subfigure}
~
\begin{subfigure}[b]{0.48\linewidth}
\centering\includegraphics[width=\textwidth]{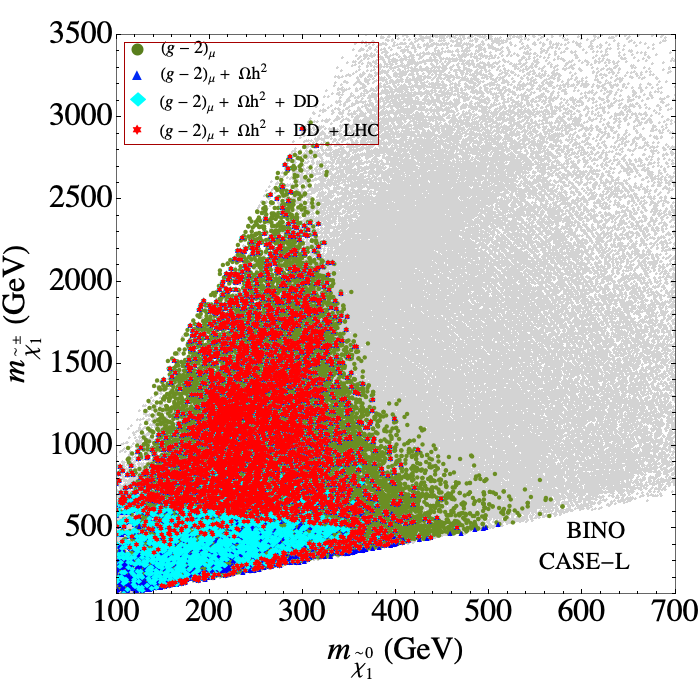}
\caption{}
\label{}
\end{subfigure}
\caption{The results of our parameter scan in the $\mneu1-\mcha1$ plane
for the $\Sl$-coannihilation case-L scenario for current (left) and
anticipated future limits (right) from \gmin2.
The color coding is as in \protect\reffi{mneu1-msmu1-caseL}.
}
\label{mneu1-mcha1-caseL}
\end{figure}

In \reffi{mneu1-msmu1-caseL} we show the results of our scan
in the $\mneu1$--$\msmu1$ plane, as before for the current \gmin2\
constraint (left) and the anticipated future constraint (right). 
The color coding of the points is the same as in \reffi{mn1-mc1-higgsino},
see the description in the beginning of \refse{sec:higgsino}.
By definition of the scenario, the points are located along the
diagonal of the plane.
The result is qualitatively similar to the analysis
in \citere{CHSold}, where the relic density constraint was used as a
direct measurement, and not only as an upper limit. However, one can observe
that a slightly larger mass difference between the $\Smu1$ and the
$\neu1$ is allowed in comparison with \citere{CHSold}. 
Taking all bounds into account we find upper limits on the LSP mass of
$\sim 500 \gev$ and $\sim 450 \gev$ for the current and
anticipated future \gmin2\ accuracy, respectively. The limits on
$\msmu1$ are about $\sim 50 - 100 \gev$ larger.

In \reffi{mneu1-mcha1-caseL} we show the results in the
$\mneu1$--$\mcha1$ plane with the same color coding as
in \reffi{mneu1-msmu1-caseL}. The results are again qualitatively in
agreement with \citere{CHSold}. However, in particular for the
future \gmin2\ constraint, slightly larger values of $\mcha1$ are
allowed. Overall we find for the light charginos mass an upper limit
of $\sim 3200 \gev$ ($\sim 2900 \gev$) for the current
(anticipated future \gmin2\ ) constraint.
Clearly visible are the LHC constraints for $\neu2$--$\chapm1$
pair production leading to three leptons and $\met$ in the final
state~\cite{Aaboud:2018jiw}.
At very low values of both $\mneu1$ and $\mcha1$ the compressed
specta searches~\cite{Aad:2019qnd} cut away points up to
$\mneu1 \lsim 150 \gev$.

\begin{figure}[htb!]
\centering
\begin{subfigure}[b]{0.48\linewidth}
\centering\includegraphics[width=\textwidth]{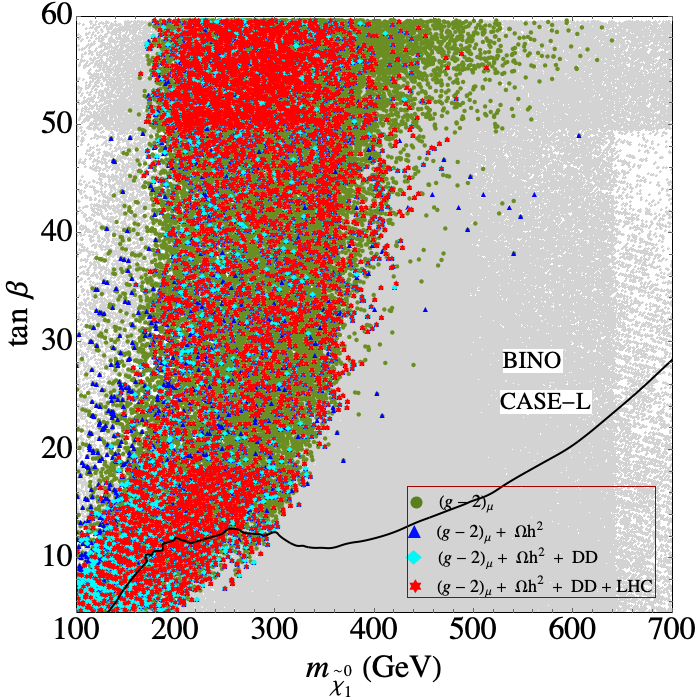}
\caption{}
\label{}
\end{subfigure}
~
\begin{subfigure}[b]{0.48\linewidth}
\centering\includegraphics[width=\textwidth]{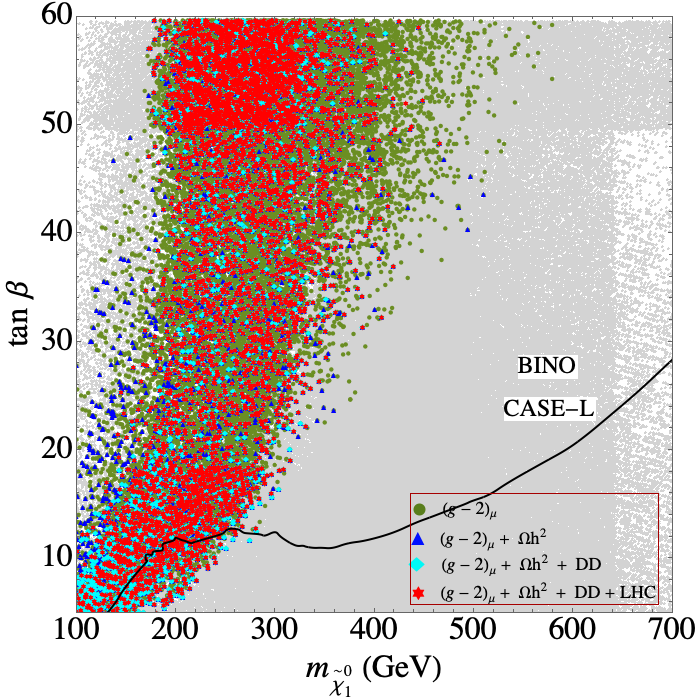}
\caption{}
\label{}
\end{subfigure}
\caption{The results of our parameter scan in the $\mneu1$--$\tb$ plane
in the  $\Sl$-coannihilation case-L scenario for current (left) and
anticipated future limits (right) from \gmin2.
The color coding is as in \protect\reffi{mneu1-msmu1-caseL}.
The black line indicates the current exclusion bounds for heavy MSSM
Higgs bosons at the LHC (see text).}
\label{mneu1-tb-caseL}
\end{figure}

The results for the $\Slpm$-coannihilation Case-L in the
$\mneu1$--$\tb$ plane are presented in \reffi{mneu1-tb-caseL}%
\footnote{
Clearly visible in the two plots are different point densities due to
indepedent samplings that were subsequently joined. However, the ranges
covered by the different searches (or colors) are clearly visible and
not affected by the different point densities.
}%
.
~The overall picture is similar to the $\cha1$-coannhiliation case shown
above in \reffi{mneu1-tb-chaco}. As above they are also qualitatively
and quantitatively similar to the results of Case-L with the relic
density taken as a direct measurement as presented in \citere{CHSold}.
Larger LSP masses are allowed for
larger $\tb$ values. On the other hand the combination of small $\mneu1$
and large $\tb$ leads to a {\it too large} contribution to
$\amu^{\rm SUSY}$ and is thus excluded. As in the other scenarios
we also show the limits from $H/A$ searches at the LHC as a solid line.
In this case for the current and anticipated future \gmin2\ limit
substantially more points passing  the \gmin2\ constraint ``survive''
below the black line(s), i.e.\ they are potential candidates for $A$-pole
annihilation. The masses reach up to $\sim 300 \gev$
and $\sim 250 \gev$, respectively. These limits are slightly
higher compared to the case with the relic density measurement taken
as a direct measurement~\cite{CHSold}.



\subsubsection{Bino DM with \boldmath{$\Slpm$}-coannihilation case-R}
\label{sec:slepR}

We now turn to our fifth scenario, bino DM with $\Slpm$-coannihilation Case-R,
where in the scan we require the ``right-handed'' sleptons to be close
in mass with the LSP. It should be kept in mind that in our notation
we do not mass-order the sleptons: for negligible mixing as it is
given for selectrons and smuons the ``left-handed'' (``right-handed'')
slepton corresponds to $\Sl_1$ ($\Sl_2$). 
As it will be seen below, in this scenario all
relevant mass scales are required to be relatively light by the
\gmin2\ constraint%
\footnote{
The scan for case-R turned out to be computationally extremely expensive,
resulting in a relatively low density of LHC tested
points. Consequently, all upper bounds on particle masses should be
taken with a relatively large uncertainty. 
}%
.

We start in \reffi{mneu1-msmu2-caseR} with the $\mneu1$--$\msmu2$ plane
with the same color coding as in \reffi{mn1-mc1-chaco}. By definition
of the scenario the points are concentrated on the diagonal.
The current (future) \gmin2\ bound yields upper limits on the LSP of
$\sim 700 (600) \gev$, as well as an upper limit on $\msmu2$ (which is close
in mass to the $\Sel2$ and $\Stau2$) of $\sim 800 (700) \gev$.
These limits remain unchanged by the inclusion of the DM relic density bound.
The limits agrees well with the ones found in the case with the relic
density measurement taken as a direct measurement~\cite{CHSold}.
Including the DD and LHC constraints, these limits reduce to
$\sim 500~(380) \gev $ for the LSP for the current
(future) \gmin2\ bounds, and correspondingly to
$\sim 550~(440) \gev$ for $\msmu2$.
The LHC constraints cut out some, but not all lower-mass points,
where the searches for compressed slepton/neutralino
spectra~\cite{Aad:2019qnd} are most relevant.
Due to the larger splitting the ``right-handed'' stau turns out to
be the NLSP in this scenario, where the uppoer bounds are found at
$\sim 500~(380) \gev$ for the current (anticipated
future) \gmin2\ bounds.

\begin{figure}[htb!]
\centering
\begin{subfigure}[b]{0.48\linewidth}
\centering\includegraphics[width=\textwidth]{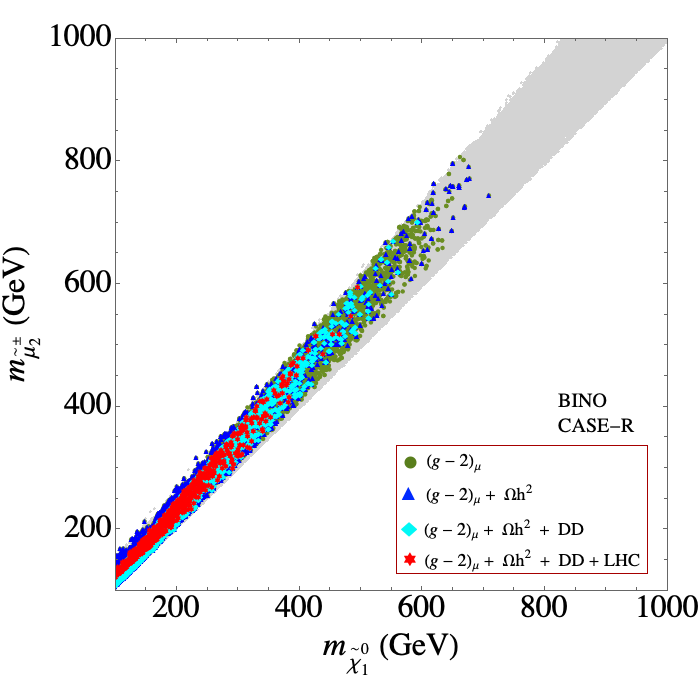}
\caption{}
\label{}
\end{subfigure}
~
\begin{subfigure}[b]{0.48\linewidth}
\centering\includegraphics[width=\textwidth]{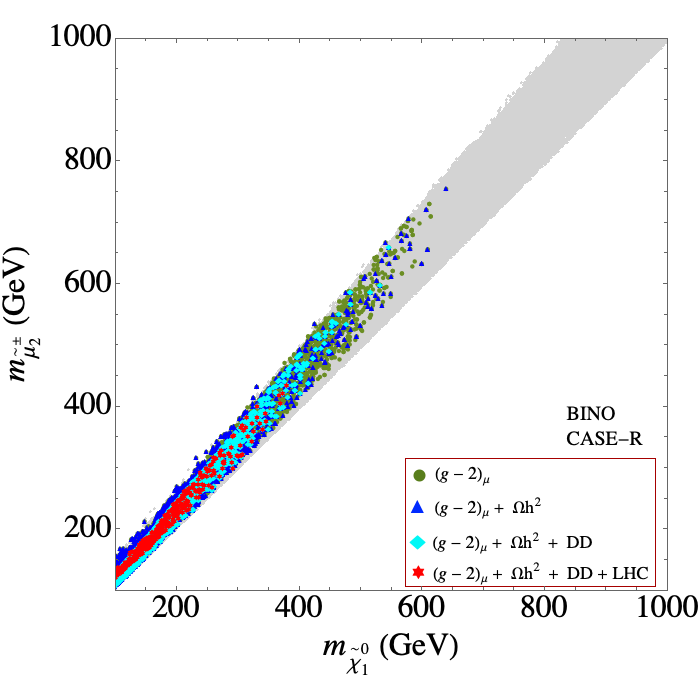}
\caption{}
\label{}
\end{subfigure}
\caption{The results of our parameter scan in the
$\mneu1$--$\msmu2$ plane
for the  $\Sl$-coannihilation case-R scenario for current (left) and
anticipated future limits (right) from \gmin2.
The color coding is as in \protect\reffi{mneu1-msmu1-caseL}.
}
\label{mneu1-msmu2-caseR}
\end{figure}

\begin{figure}[htb!]
	\vspace{2em}
\centering
\begin{subfigure}[b]{0.48\linewidth}
\centering\includegraphics[width=\textwidth]{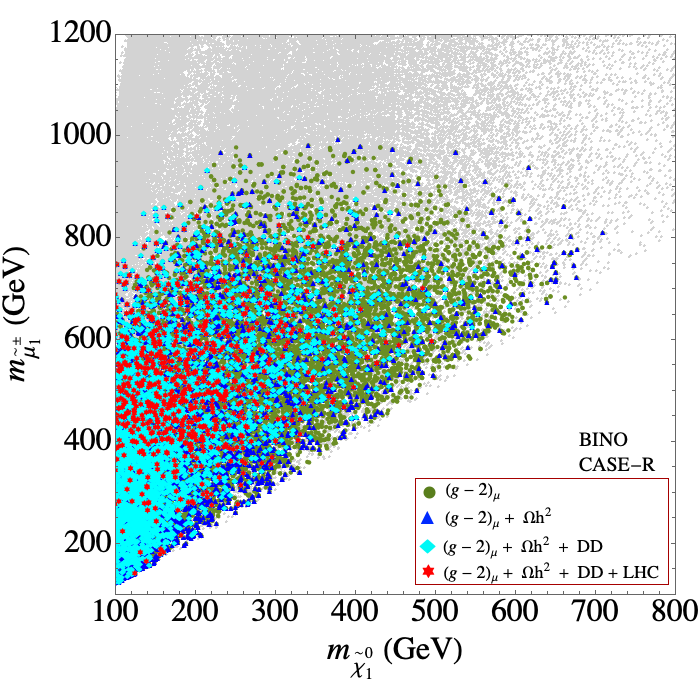}
\caption{}
\label{}
\end{subfigure}
~
\begin{subfigure}[b]{0.48\linewidth}
\centering\includegraphics[width=\textwidth]{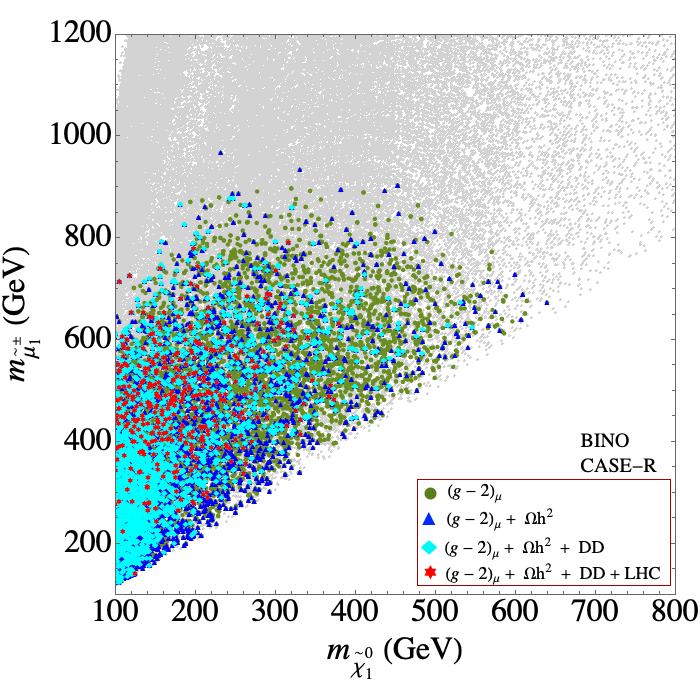}
\caption{}
\label{}
\end{subfigure}
\caption{The results of our parameter scan in the
$\mneu1$--$\msmu1$ plane
for the  $\Sl$-coannihilation case-R scenario for current (left) and
anticipated future limits (right) from \gmin2.
The color coding is as in \protect\reffi{mneu1-msmu2-caseR}.
}
\label{mneu1-msmu1-caseR}
\end{figure}

The distribution of the heavier slepton is displayed in the
$\mneu1$--$\msmu1$ plane in \reffi{mneu1-msmu1-caseR}. As before, the
results agree well with the ones found in \citere{CHSold}. 
Although the ``left-handed'' sleptons are allowed to be much heavier,
the \gmin2\ constraint imposes an upper limit of $\sim 950~(800) \gev$
in the case for the current (future) \gmin2\ precision.
Taking into account the CDM and LHC constraints we find upper limits for
$\msmu1$ of $\sim 850 \gev$ and $\sim 800 \gev$ for the
current and anticipated future \gmin2\ constraint, respectively.

In \reffi{mneu1-mcha1-caseR} we show the results in the
$\mneu1$--$\mcha1$ plane with the same color coding as
in \reffi{mneu1-msmu2-caseR}.
The results are again in qualitative agreement with the case of the DM
relic density taken as a direct measurement~\cite{CHSold}. 
As in the Case-L the \gmin2\ limits on
$\mneu1$ become slightly stronger for larger chargino masses.
The upper limits on the
chargino mass, however, are substantially stronger as in the Case-L.
Taking all constraints into account, they are
reached at $\sim 1540 \gev$ for the current and $\sim 1350 \gev$
for the anticipated future precision in $\amu$. 

\begin{figure}[htb!]
\centering
\begin{subfigure}[b]{0.48\linewidth}
\centering\includegraphics[width=\textwidth]{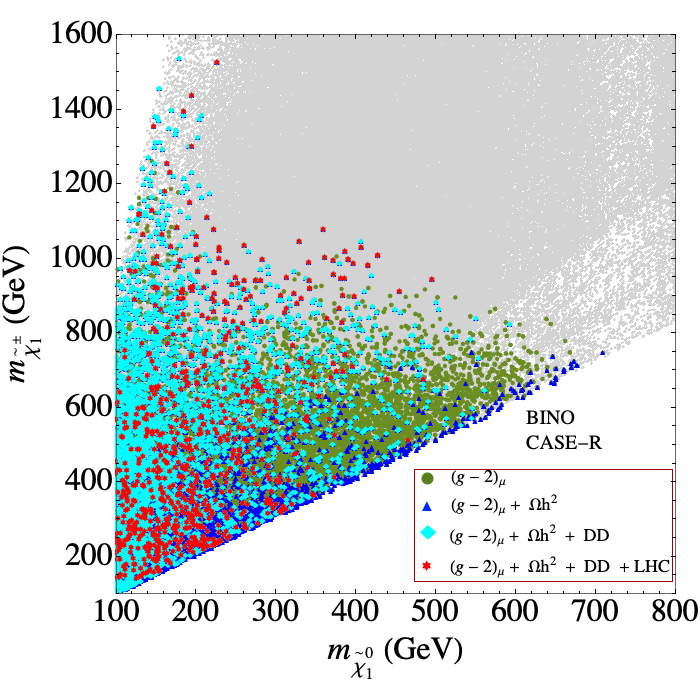}
\caption{}
\label{}
\end{subfigure}
~
\begin{subfigure}[b]{0.48\linewidth}
\centering\includegraphics[width=\textwidth]{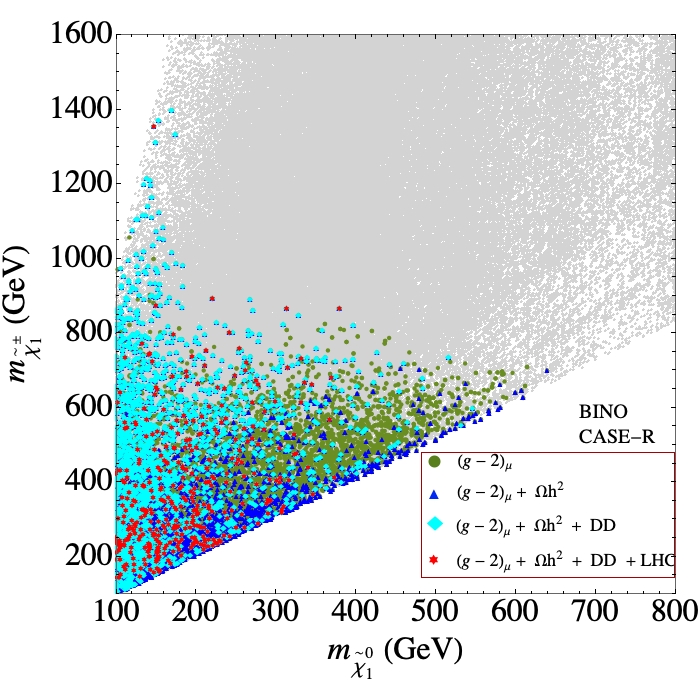}
\caption{}
\label{}
\end{subfigure}
\caption{The results of our parameter scan in the $\mneu1-\mcha1$ plane
for the $\Sl$-coannihilation case-R scenario for current (left) and
anticipated future limits (right) from \gmin2. 
The color coding is as in \protect\reffi{mneu1-msmu2-caseR}.
}
\label{mneu1-mcha1-caseR}
\end{figure}

\begin{figure}[htb!]
\centering
\begin{subfigure}[b]{0.48\linewidth}
\centering\includegraphics[width=\textwidth]{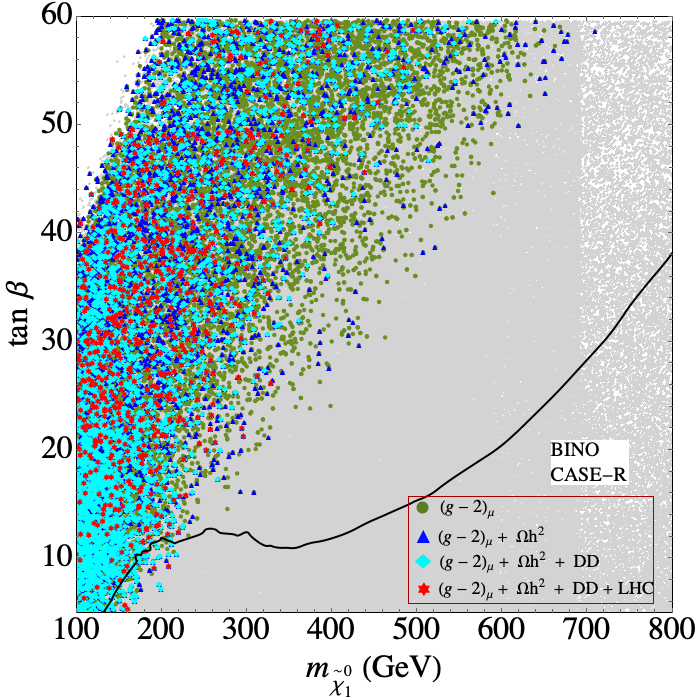}
\caption{}
\label{}
\end{subfigure}
~
\begin{subfigure}[b]{0.48\linewidth}
\centering\includegraphics[width=\textwidth]{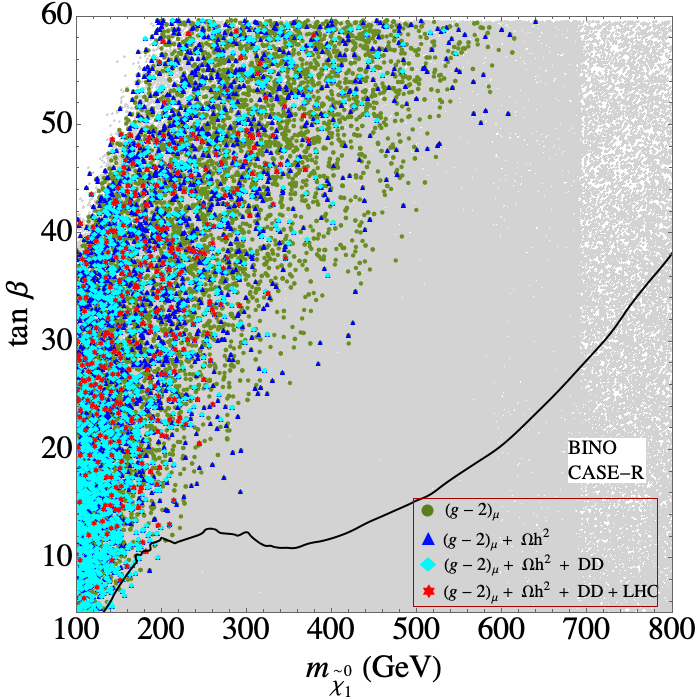}
\caption{}
\label{}
\end{subfigure}
\caption{The results of our parameter scan in the $\mneu1$--$\tb$ plane
in the  $\Sl$-coannihilation case-R scenario for current (left) and
anticipated future limits (right) from \gmin2.
The color coding is as in \protect\reffi{mneu1-msmu2-caseR}.
The black line indicates the current exclusion bounds for heavy MSSM
Higgs bosons at the LHC (see text). 
}
\label{mneu1-tb-caseR}
\end{figure}

We finish our analysis of the $\Slpm$-coannihilation Case-R  with the
results in the $\mneu1$--$\tb$ plane, presented in \reffi{mneu1-tb-caseR}. The
overall picture is similar to the previous cases shown
above in \reffis{mneu1-tb-chaco}, \ref{mneu1-tb-caseL}, and also
qualitatively in agreement with the results found in \citere{CHSold}.
Larger LSP masses are allowed for larger $\tb$ values. On the other
hand the combination of small $\mneu1$ and very large $\tb$ values,
$\tb \gsim 40$ leads to stau masses below the LSP mass, which we
exclude for the CDM constraints.
The LHC searches mainly affect parameter points with $\tb \lsim 20$. 
Larger $\tb$ values induce a larger mixing in the third
slepton generation, enhancing the probability for charginos to
decay via staus and thus evading the LHC constraints. As before 
we also show the limits from $H/A$ searches at the LHC, where we set
(as above) $\mneu1 = \MA/2$, i.e.\ roughly to the requirement for $A$-pole
annihilation, where points above the black lines are experimentally excluded. 
Comparing Case-R and Case-L, here for the current \gmin2\ limit
substantially less points are passing the current \gmin2\ constraint
below the black line, i.e.\ are potential candidates for $A$-pole
annihilation. The masses reach only up to $\sim 200 \gev$.
With the anticipated future \gmin2\ limit hardly any point survives,
leaving $A$-pole annihilation as a quite remote
possibility with a strict upper bound on $\mneu1$.


\subsection{Lowest and highest mass points}
\label{sec:spectra}

In this section we present some sample spectra for the five cases
discussed in the previous subsections. For each case, higgsino DM, wino
DM, bino/wino DM with $\cha1$-coannihilation, $\Slpm$-coannihilation
Case-L and Case-R, we 
present three parameter points that are in agreement with all
constraints (red points): the lowest LSP mass, the highest LSP
with current \gmin2\ constraints, as well as the highest LSP mass
with the anticipated future \gmin2\ constraint.
They will be labeled as ``H1, H2, H3'', ``W1, W2, W3'', "C1, C2, C3",
"L1, L2, L3", "R1, R2, R3" for higgsino DM, wino DM, bino/wino DM with 
$\chapm1$-coannihilation, $\Slpm$-coannihilation Case-L and
Case-R, respectively.
While the points are obtained from "random sampling",
nevertheless they give an idea of the mass spectra realized in the
various scenarios.
In particular, the highest mass points give a clear indication on the
upper limits of the NLSP mass.
It should be noted that the $\ssi$ values given below are scaled
with a factor of ($\Omega_{\tilde \chi} h^2$/0.118) to account for the
lower relic density.

In \refta{bptab1} we show the $3$ parameter points ("H1, H2, H3") from
higgsino DM scenario, which are
defined by the six scan parameters: $M_1$, $M_2$, $\mu$, $\tb$ and the
two slepton mass parameters, $\msl{L}$ and $\msl{R}$ (corresponding
roughly to $m_{\tilde e_1, \tilde \mu_1}$ and
$m_{\tilde e_2, \tilde \mu_2}$, respectively).
Together with the
masses and relevant $\br$s we also show the values of the
DM observables and $\amu^{\rm SUSY}$.
By definition of the scenario we have $\mu < M_1, M_2$ and
$\mneu1 \sim \mneu2 \sim \mcha1$ with the $\cha1$ as NLSP.
We find in all three points $M_2 \ll M_1$, which can be understood
from the \gmin2\ constraint. For (relatively) small $\mu$ the
chargino-sneutrino contribution, \refeq{amuchar}, is $\propto 1/M_2$ and
thus becomes large for smaller $M_2$. Conversely, the neutralino-slepton
contribution, \refeq{amuslep}, is $\propto M_1$ and thus becomes larger
for larger $M_1$.
As anticipated, the relic density is found to be quite low.
For all of the three points, the decays of $\chapm1$ and $\neu2$
to sleptons are not kinematically accessible.
Therefore they mainly decay via off-shell sleptons and/or gauge-bosons
to a pair of SM particles and the LSP. 
The sleptons represented by $\sle1$ and $\sle2$ decay to
chargino/neutralino and the corresponding SM particle.
The $\sle2$ is mostly ``right-handed'' and thus decays preferably
to a bino and an electron. However, this mode is kinematically open only
for H3, leading to the observed $\sle2$ decay pattern for H1 and H2.

\begin{table}[!htb]
{\small
\begin{center}
\hspace{-1 cm}
\begin{tabular}[t]{|c||c|c|c||l||c|c|c|}
\hline
Sample points          & H1   & H2    & H3     &  Sample points                      & H1         & H2   & H3   \\
\hline
\hline
$M_1$                  & 1005  & 3414 &  3397  &$\br(\neu2 \ra \neu1 \gamma$                        & 0.5   &  5   & 2.4     \\
$M_2$                  & 478  & 1165 &  1007  & $\phantom{\br(\neu2 } \ra \neu1 q \bar q$           & 51    & 42.5 & 41.4    \\  
$\mu$                  & 124  &  472  &  454   & $\phantom{\br(\neu2 } \ra \neu1 l \bar l$          & 6     & 5.8  &5.6      \\
$\tb$                  &  56  &  51.6   &52.8  & $\phantom{\br(\neu2 } \ra \neu1 \tau \bar\tau$     & 3     & 2.8  & 2.4        \\
$\msl{L}$              &  599 &  496  &  520    &$\phantom{\br(\neu2 } \ra \neu1 \nu \bar\nu$      &18     & 17.7 & 17         \\
$\msl{R}$              &  744 &  3039 &  4402   &$\phantom{\br(\neu2 } \ra \chapm1 q \bar q$       & 14    & 17.4 & 21         \\
$\mneu1$               &  118.6 &  475 &  455    &$\phantom{\br(\neu2 } \ra \chapm1 \nu_l l$       & 4.8   & 7    & 8          \\
$\mneu2$               & 133   & 482    &463     &$\phantom{\br(\neu2 } \ra \chapm1 \nu_\tau \tau$) & 2     & 1.7  & 2.4        \\
\cline{5-8}
$\mneu3 \sim \mcha2$   &   515 &1200     &1043   &$\br(\chapm1 \ra \neu1 \tau \nu_\tau$      &7.8       & 0.1  & 0.4       \\
$\mneu4 $              & 993   &  3390    & 3372 &$\phantom{\br(\chapm1 } \ra \neu1 l \nu_l$ & 25.6     & 38   & 37       \\
                       &       &          &      &$\phantom{\br(\chapm1} \ra \neu1  q \bar q'$) &   66.5   & 61.5 & 62.4     \\                       
\cline{5-8}
$\mcha1$               &  124  & 477   &  457.6    &$\br(\sle1 \ra \neu1 l$             &5.7    & 57  & 51\\
$m_{\tilde e_1,
\tilde \mu_1}$         &  601  & 498   & 522    &  $\phantom{\br(\sle1} \ra \neu2 l$   & 1.9   & 4   &5    \\
$m_{\tilde e_2,
\tilde \mu_2}$         & 745  &  3040  &4401  & $\phantom{\br(\sle1 } \ra \neu3 l$ &  29    &  -   & -   \\        
$m_{\tilde \tau_1}$    & 600   &498    & 522   & $\phantom{\br(\sle1 } \ra \chapm1 \nu_l$ &2.8   &39   & 44 \\  
$m_{\tilde \tau_2}$    & 746   &3040    & 4402    & $\phantom{\br(\sle1 } \ra \chapm2 \nu_l$) &60   & -  & - \\
\cline{5-8}
$m_{\tilde \nu}$        & 596  &492  &   516  &$\br(\sle2 \ra \neu1 l $     & 59 &64  & - \\   
$\Omega_{\tilde \chi} h^2$  & 0.003  & 0.036 & 0.024  &$\phantom{\br(\sle2 } \ra \neu2 l $ & 37  &34   & -\\
$a_\mu^{\rm SUSY} \tenp{10}$& 37.5  &14.1  & 16.5  & $\phantom{\br(\sle2 } \ra \neu3 l $ &  3    & 1     &  -  \\
$\ssi \tenp{10}$            & 1  & 3.5  &3.5     &$\phantom{\br(\sle2 } \ra \neu4 l $) & -     & -      & 99.9   \\
\hline
\end{tabular}
 \end{center}
}
\caption{The masses (in $\gev$) and relevant $\br$s (\%) of three points
from the higgsino DM scenario,
corresponding to the lowest LSP mass, the highest LSP mass
with current \gmin2\ constraints, as well as the highest LSP mass
with the anticipated future \gmin2\ constraint.
Here $\Sl\,(l)$ refers to $\tilde e\, (e)$ and $\tilde \mu\, (\mu)$
together. $\nu$ denotes
$\nu_e$, $\nu_\mu$ and $\nu_\tau$ together.
Only $\br$s above $0.1$ \% are shown. The values of $\gmin2$ and DM
observables are also shown.
$\ssi$ is scaled with a factor of ($\Omega_{\tilde \chi} h^2$/0.118) and
given in the units of~pb.
}
\label{bptab1}
\end{table}

\medskip
In \refta{bptab2} we show the $3$ parameter points ("W1, W2, W3") from
wino DM scenario, defined in terms of the same set of input parameters
as the higgsino DM scenario.
Together with the
masses and relevant $\br$s we also show the values of the
DM observables and $\amu^{\rm SUSY}$.
By definition of the scenario we have $M_2 < M_1, \mu$ and
$\mneu1 \sim \mcha1$, i.e.\ the $\cha1$ being the NLSP decaying
dominantly as $\cha1 \to \neu1 \pi^\pm$, but also the decay to the LSP
and $e \nu_e$ or $\mu \nu_\mu$ occurs (either with a soft pion or a soft
charged lepton).
As anticipated, the relic density is found to be quite low.
The second lightest neutralino is found substantially heavier than the
LSP, with a variety of decay channels to be open, diluting possible
signals. 
The sleptons represented by $\sle1$ and $\sle2$ decay to
chargino/neutralino and the corresponding SM particle with relevant BRs
to charged leptons, which will be crucial for their detection.
The $\sle2$ is mostly ``right-handed'' and thus decays preferably
to a bino and an electron (i.e.\ the neutralino in the decay is
determined by the mass ordering of $\mu$ and $M_1$).

\begin{table}[!htb]
{\small
\begin{center}
\hspace{-1 cm}
\begin{tabular}[t]{|c||c|c|c||l||c|c|c|}
\hline
Sample points          & W1   & W2    & W3     &  Sample points                      & W1         & W2   & W3   \\
\hline
\hline
$M_1$                  & 948  &747  & 1061     &$\br(\neu2 \ra \tilde l_1  l$                    & -   &  32.8   & 3.56    \\
$M_2$                  & 110  &608  & 523      & $\phantom{\br(\neu2 } \ra \Stau1 \tau$          & -   &  19.4   & 18.6    \\  
$\mu$                  & 313  & 1122  &949     & $\phantom{\br(\neu2 } \ra \tilde \nu \nu $      & -   &  44.4   & 0.12      \\
$\tb$                  & 20.3 &  56.5 &57.4    & $\phantom{\br(\neu2 } \ra \chapm1 W$           & 70  &  3.2    & 52        \\
$\msl{L}$              & 557  &  679   &656     & $\phantom{\br(\neu2 } \ra \neu1 Z$            & 27    &  -      & 1.8        \\
$\msl{R}$              & 356 &  1744  &1559    & $\phantom{\br(\neu2 } \ra \neu1 h)$            & 3    &  -      & 24    \\
$\mneu1$               & 100 &  602.8 &517.7   &                                                 &     &        &        \\
$\mneu2$               & 320  & 745   &948     &                                                 &     &        &        \\
\cline{5-8}
$\mneu3 \sim \mcha2$   & 328 & 1124   &951     &$\br(\chapm1 \ra \neu1 q \bar q'$          &  67.4   & 73  & 74       \\
$\mneu4 $              & 950  & 1132  & 1070   &$\phantom{\br(\chapm1 } \ra \neu1 e \nu_e$ &   22.9 & 27  & 26    \\
                       &      &       &        &$\phantom{\br(\chapm1 } \ra \neu1 \mu \nu_\mu)$   & 9.7    &   -     & -       \\
\cline{5-8}
$\mcha1$               & 100.4 & 603   & 517.8 &$\br(\sle1 \ra \neu1 l$                   & 32.5   & 33  & 33.4\\
$m_{\tilde e_1,
\tilde \mu_1}$         & 559  &680    & 658    & $\phantom{\br(\sle1} \ra \neu3 l$                   & 1.4   &  -      & -       \\
$m_{\tilde e_2,
\tilde \mu_2}$         & 359  & 1745  & 1560  &$\phantom{\br(\sle1} \ra \chapm1 \nu_l$   &61.6   & 67   &66.6    \\
$m_{\tilde \tau_1}$    & 358  & 676   &  653  &    $\phantom{\br(\sle1} \ra \chapm2 \nu_l)$  & 4.2    &   -     &   -     \\
$m_{\tilde \tau_2}$    & 560   & 1746 & 1561  &                                                 &     &        &        \\ 
\cline{5-8}
$m_{\tilde \nu}$        & 554   &676  & 653  &$\br(\sle2 \ra \neu1 l $                    & 30.4  & -  & - \\   
$\Omega_{\tilde \chi} h^2$  & 3.6 $\times 10^{-4}$  & 0.014  & 0.009  & $\phantom{\br(\sle2}\ra \neu2 l $  & 16.4     & 99.6       & 13       \\
$a_\mu^{\rm SUSY} \tenp{10}$&16.2   & 13.7  & 16.5  &$\phantom{\br(\sle2 } \ra \neu3 l$ & 53  & -   & - \\
$\ssi \tenp{10}$            & 0.12  & 1.6  & 1.48  &$\phantom{\br(\sle2 } \ra \neu4 l)$ & -    & 0.3   & 87       \\
\hline
\end{tabular}
 \end{center}
}
\caption{The masses (in $\gev$) and relevant $\br$s (\%) of three points
from the wino DM scenario,
corresponding to the lowest LSP mass, the highest LSP mass
with current \gmin2\ constraints, as well as the highest LSP mass
with the anticipated future \gmin2\ constraint. The notation and units
used are as in \protect\refta{bptab1}. $\tilde\nu \nu$ refers to
all three generations (which in our sampling are mass degenerate).
Only $\br$s above $0.1$ \% are shown.  
}
\label{bptab2}
\end{table}

\medskip

In \refta{bptab3} we show the $3$ parameter points ("C1, C2, C3") from
$\chapm1$-coannihilation scenario, with the same definitions and
notations as for the previous tables.
We find that the $\tilde \tau$ is the NLSP in these three cases, and the 
combined contribution from $\tilde \tau$-coannihilation together
with $\chapm1$-coannihilation brings the relic density
to the ballpark value. 
For all of the three points, the decays of $\chapm1$ and $\neu2$
to first two generations of sleptons are not kinematically
accessible or strongly phase space suppressed.
Therefore they decay with a very large BR to third generation charged
sleptons and sneutrinos. This makes them effectively invisible to
the LHC searches looking for 
electrons and muons in the signal. LHC analyses designed to
specifically look for $\tau$-rich final states can prove beneficial
to constrain these points, which are much less powerful, as
discussed above. We refrain from showing results for $\Stau{}$
decays, since the corresponding
dedicated searches turn out to be weaker (and thus not effective) than
other applicable searches.

\begin{table}[!htb]
{\small
\begin{center}
\hspace{-1 cm}
\begin{tabular}[t]{|c||c|c|c||l||c|c|c|}
\hline
Sample points          & C1   & C2    & C3     &  Sample points                      & C1         & C2   & C3   \\
\hline
\hline
$M_1$                  & 138  &592  & 528    &$\br(\neu2 \ra \tilde \tau_1 \tau$  & 95.6      & 100 &100 \\
$M_2$                  & 151  &640  & 556    &$\phantom{\br(\neu2} \ra \tilde{\nu} \nu$)   & 4.3   &  -   & -\\  
$\mu$                  & 1113 & 1094  & 1032 &                                     &      &     &  \\
\cline{5-8}
$\tb$                  & 6.3 & 57    &  55   &$\br(\chapm1 \ra \tilde \tau_1 \nu_{\tau}$ & 96.3  & 100 & 100 \\ 
$\msl{L} = \msl{R}$    & 167 & 678    & 616  &$\phantom{\br(\chapm1} \ra l \tilde \nu_l$         & 3     &   -   &-    \\
$\mneu1$               & 134 & 583    & 520  &$\phantom{\br(\chapm1} \ra \tau \tilde \nu_\tau$)   & 0.6   &   -   & -   \\
$\mneu2$               & 158  &663    & 576  &                                            &       &      &    \\
$\mneu3$               & 1123 &1100   & 1039 &                                            &       &      &    \\
\cline{5-8}
$\mneu4 \sim \mcha2$   & 1126& 1107  & 1045  &$\br(\sle1 \ra \neu1 l$             & 32   &  74 & 33\\
$\mcha1$               & 158 & 662   & 578    &$\phantom{\br(\sle1} \ra \neu2 l$  & 23   &  9 & 23   \\
$m_{\tilde e_1,
\tilde \mu_1}$         & 173& 680   & 618    &$\phantom{\br(\sle1 } \ra \chapm1 \nu_l$) & 45   & 17 & 44  \\
$m_{\tilde e_2,
\tilde \mu_2}$         & 173& 680   & 618    &    &      &      &    \\       
$m_{\tilde \tau_1}$    & 134  & 584   & 523   &                                      &      &      &    \\       
$m_{\tilde \tau_2}$    & 204  & 764   & 700   &                                      &      &      &     \\
\cline{5-8}
$m_{\tilde \nu}$        & 155   & 675 & 613         &$\br(\sle2 \ra \neu1 l) $
& 99.9 & 99.9 & 99.9\\   
$\Omega_{\tilde \chi} h^2$  &0.02  & 0.07 & 0.078   &          &   &   &  \\
$a_\mu^{\rm SUSY} \tenp{10}$&24   & 14.3  & 16.5    &                                      &      &      &    \\
$\ssi \tenp{10}$            & 0.083  & 1.09  & 1.18 &                                      &      &      &    \\
\hline
\end{tabular}
 \end{center}
}
\caption{The masses (in $\gev$) and relevant $\br$s (\%) of three points
from $\chapm1$-coannihilation
scenario corresponding to the lowest LSP mass, the highest LSP mass
with current \gmin2\ constraints, as well as the highest LSP mass
with the anticipated future \gmin2\ constraint.
The notation, definitions and units are as in \refta{bptab2}. 
Only $\br$s above $0.1$ \% are shown. 
}
\label{bptab3}
\end{table}

\medskip

In \refta{bptab4} we show three parameter points ("L1, L2, L3") taken
from $\Slpm$-coannihilation scenario Case-L, defined in the same way as
in the $\chapm1$-coannihilation case.
The character of the $\neu2$ and $\cha1$ depend on the mass ordering of
$\mu$ and $M_2$ and are thus somewhat random. This leads to a large
variation in the various BRs of these two particles, possibly diluting
any signal involving a certain SM particle, such as a charged lepton.
The selectrons or smuons, on the other hand, decay to the LSP and the
corresponding SM lepton, which tend to be soft in this case,
making way for compressed spectra searches at the LHC.

\begin{table}[!htb]
{\small
\begin{center}
\hspace{-1 cm}
\begin{tabular}[t]{|c||c|c|c||l||c|c|c|}
\hline
Sample points          & L1   & L2   & L3    &  Sample points          & L1         & L2   & L3   \\ 

\hline
\hline
$M_1$                  & 102  & 522 & 474          &$\br(\neu2  \ra \Sl_1 l$     & 22   & 19 & 26\\
$M_2$                  & 1008  & 903 & 504        &$\phantom{\br(\neu2 }\ra \Sl_2 l$      & 1    & - & - \\
$\mu$                  & 675 & 896 &  1262        &$\phantom{\br(\neu2 } \ra \tilde \tau_1 \tau$ & 14.4 & 45 & 25.6 \\  
$\tb$                  & 6.8  & 55.6  & 48.6      &$\phantom{\br(\neu2 } \ra \tilde \tau_2 \tau$ &  1.6    & 4.4   &    \\
$\msl{L}$              & 118 & 601 & 483          &$\phantom{\br(\neu2 } \ra \neu1 h$            &  35    & 7.6    & -\\  
$\msl{R}$               & 407 & 606  & 1024       &$\phantom{\br(\neu2 } \ra \neu1 Z$           & 11     & 0.5    & -\\  
$\mneu1$               & 97 & 513  & 466         &$\phantom{\br(\neu2 } \ra \tilde \nu \nu)$     & 15 & 23 &48.6 \\ 
$\mneu2$               & 675  & 860 &  525       &                                     &      &     & \\  
$\mneu3$               & 686  & 904 &  1269      &                                     &      &     & \\  
\cline{5-8}
$\mneu4 \sim \mcha2$   & 1045  & 975 & 1272      &$\br(\chapm1 \ra \tilde \nu_{l} l$    & 21.6 & 19 & 33.4\\  
$\mcha1$               & 674 & 860 & 525         &$\phantom{\br(\chapm1 } \ra \tilde \nu_{\tau} \tau$& 14 & 24 & 17\\  
$m_{\tilde e_1,
\tilde \mu_1}$         & 126  & 602 & 485        &$\phantom{\br(\chapm1 } \ra \Sl_1 \nu_l$         &  11  & 16 & 25\\   
$m_{\tilde e_2,
\tilde \mu_2}$         & 407  & 608 & 1025       &$\phantom{\br(\chapm1 } \ra \tilde \tau_1 \nu_\tau$ &  6  & 33    & 24.5 \\  
$m_{\tilde \tau_1}$    & 124  & 519 &  468       &$\phantom{\br(\chapm1 } \ra \tilde \tau_2 \nu_\tau$ & 1  & 0.3   & -\\  
$m_{\tilde \tau_2}$    & 410 & 681 & 1033        &$\phantom{\br(\chapm1 } \ra W \neu1)$               & 46 & 7.5  & - \\  
$m_{\tilde \nu}$        &100  &597  & 479        &                                     &      &     & \\  
\cline{5-8}
$\Omega_{\tilde \chi} h^2$  & 0.036 & 0.121 & 0.109 &$\br(\sle1 \ra \neu1 l)$ & 100     & 100    &100 \\
\cline{5-8}
$a_\mu^{\rm SUSY} \tenp{10}$& 32.87 & 14 & 17.3       &$\br(\sle2 \ra \neu1 l) $         & 100    & 100    &99.99 \\  
$\ssi \tenp{10}$            & 0.25 & 2.8  & 0.45       &                               &      &     & \\  
\hline
\end{tabular}
 \end{center}
}
\caption{The masses (in $\gev$) and relevant $\br$s (\%) of three points
from $\Slpm$-coannihilation
scenario Case-L corresponding to the lowest LSP mass, the highest LSP mass
with current \gmin2\ constraints, as well as the highest LSP mass
with the anticipated future \gmin2\ constraint.
The notation, definitions and units are as in \refta{bptab1}. 
Only $\br$s above $0.1$ \% are shown.
}
\label{bptab4}
\end{table}

\medskip

The masses, $\br$s and values of the $\gmin2$ and DM observables
of the parameter point for the Case-R  ("R1, R2, R3'') are
shown in \refta{bptab5}, defined in the same way as
in the $\chapm1$-coannihilation case.
As in the case-L the character of the $\neu2$ and $\cha1$ depend on the
mass ordering of $\mu$ and $M_2$ and are thus somewhat random.
In particular for the two high-mass points R2 and R3 the two mass
parameters are relatively close, leading to larger mixings for these two
states. This in turn leads to a larger
variation in the various BRs of these two particles, possibly diluting
any signal involving a certain SM particle, such as a charged lepton.
The selectrons or smuons, on the other hand, decay preferably to the LSP
and the corresponding SM lepton. As in case-L these tend to be soft, 
making way for compressed spectra searches at the LHC.

\begin{table}[!htb]
{\small
\begin{center}
\hspace{-1 cm}
\begin{tabular}[t]{|c||c|c|c||l||c|c|c|}
\hline
Sample points          & R1   & R2    & R3     &  Sample points               & R1         & R2   & R3   \\ 

\hline
\hline
$M_1$                  &103  & 504  & 385      & $\br(\neu2 \ra \Sl_2 l$ & 1.8 & 0.4 & - \\
$M_2$                  &212  & 1169  & 877     & $\phantom{\br(\neu2} \ra \Sl_1 l$ & -  & 4   & 24 \\
$\mu$                  &971  & 958  &  976     &$\phantom{\br(\neu2} \ra \tilde \tau_2 \tau$ & 97   & 61 & 32 \\                   
$\tb$                  &6.6  & 59  & 59        &$\phantom{\br(\neu2} \ra \tilde \tau_1 \tau$ & - & 18.4 & 7 \\                   
$\msl{L}$              &272     & 603   & 638     &$\phantom{\br(\neu2} \ra \neu1 h$           & 0.9   & 11.5 & 2 \\    
$\msl{R}$              &112     & 598   & 438     &$\phantom{\br(\neu2} \ra \neu1 Z$          & 0.2 &  1   & 0.3\\   
$\mneu1$               &100  & 496   & 380     &$\phantom{\br(\neu2} \ra \tilde \nu \nu)$    & -   & 3    & 34.5   \\
\cline{5-8}
$\mneu2$               &222 & 953  & 877    &$\br(\chapm1 \ra \Sl_1 \nu_l$                       &-     & 2   & 22.4 \\           
$\mneu3$               &982 & 966  & 982    &$\phantom{\br(\chapm1} \ra \tilde \tau_2 \nu_{\tau}$ & 96  & 37  & 30 \\
$\mneu4 \sim \mcha2$   &986 & 1209 & 1011   &$\phantom{\br(\chapm1} \ra \tilde \tau_1 \nu_{\tau}$ & -  &  7 & 4 \\ 
$\mcha1$               &222 & 952 & 877     &$\phantom{\br(\chapm1} \ra \tilde \nu_l l$ &   -   &4 & 24 \\ 
$m_{\tilde e_1,\tilde \mu_1}$&276 & 605  & 639 &$\phantom{\br(\chapm1} \ra \tilde \nu_{\tau} \tau$ & -  &38  &17 \\ 
$m_{\tilde e_2,
\tilde \mu_2}$         &120 & 599 & 440    & $\phantom{\br(\chapm1} \ra W \neu1$) & 4  & 12  & 2.4\\
\cline{5-8}
$m_{\tilde \tau_2}$    &112   & 501  & 382 &$\br(\sle1 \ra \neu1 l)$  &39      & 100    & 100 \\                
$m_{\tilde \tau_1}$    & 279  & 689  & 676 &$\phantom{\br(\sle1} \ra \neu2 l$    & 21      &  -   & - \\   
$m_{\tilde \nu}$        &265 & 600  & 634  &$\phantom{\br(\sle1} \ra \chapm1 \nu_l)$  &40      & -   & - \\  
$\Omega_{\tilde \chi} h^2$  &0.115 & 0.08  & 0.036 &                       &      &     &    \\
\cline{5-8}
$a_\mu^{\rm SUSY} \tenp{10}$&16.8 & 13  & 15.9   &$\br(\sle2 \ra \neu1 l)$  &100      & 100   & 100\\  
$\ssi \tenp{10}$            &0.35 & 1.24 & 0.27  &                                      &      &     &    \\  
\hline
\end{tabular}
 \end{center}
}
\caption{The masses (in $\gev$) and relevant $\br$s (\%) of three points
from $\Slpm$-coannihilation
scenario Case-R corresponding to the lowest LSP mass, the highest LSP mass
with current \gmin2\ constraints, as well as the highest LSP mass
with the anticipated future \gmin2\ constraint.
The notation, definitions and units are as in \refta{bptab1}. 
Only $\br$s above $0.1$ \% are shown. 
}
\label{bptab5}
\end{table}

\clearpage

\section{Prospects for future colliders}
\label{sec:future}

In this section we briefly discuss the prospects of the direct detection of
the (relatively light) EW particles at the approved HL-LHC, the
hypothetical upgrade to the HE-LHC, the potential future FCC-hh, and at a
possible future $e^+e^-$ collider such as
ILC~\cite{ILC-TDR,LCreport} or CLIC~\cite{CLIC,LCreport}.  
We concentrate on the compressed spectra searches, relevant for
higgsino DM, wino DM and bino/wino DM with
$\cha{1}$-coannihilation. Results for the future prospects for slepton
coannihilation (although with the relic DM density taken as
a direct measurement) can be found in \citere{CHSold}. 

\begin{figure}[htb!]
  \centering
  \includegraphics[width=0.8\textwidth]{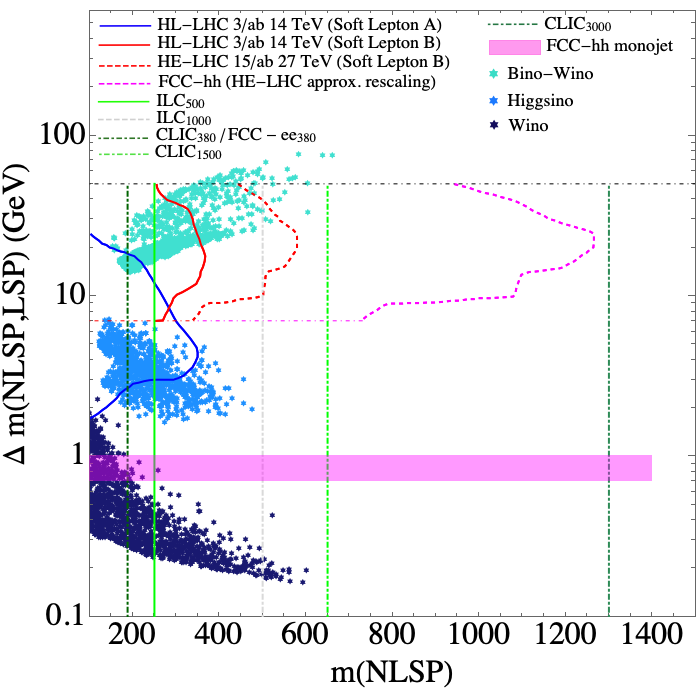}
  \caption{$\mcha{1}$--$\De m$ plane 
  with anticipated limits from compressed spectra searches at the
  HL-LHC, the HE-LHC, FCC-hh and ILC500,
  ILC1000, CLIC380, CLIC1500, CLIC3000 (see text), original taken
  from \protect\citere{Strategy:2019vxc}. Not included are disappearing
  track searches.  
Shown in blue, dark blue, turquoise are the points surviving all current
  constraints in the case of higgsino DM, wino DM and bino/wino DM with
  $\cha1$-coannihilation, respectively.
}
\label{fig:future}
\end{figure}

In \reffi{fig:future} we present our results in the $\mcha1$--$\De m$ plane
(with $\De m := \mcha1 - \mneu1$), which was presented (in its original
form, i.e.\ down to $\De m = 0.7 \gev$)
in \citere{Strategy:2019vxc} for the higgsino DM case, but also directly
valid for the wino DM case~\cite{Berggren:2020tle} (see the discussion
below). Shown are the following anticipated limits for compressed
spectra searches%
\footnote{
Very recent analysis using vector boson fusion processes~\cite{Natalia:2021ssb}
or a future muon collider~\cite{Capdevilla:2021fmj} for compressed
(higgsino) spectra have not been taken into account.
}%
:

\begin{itemize}

\item HL-LHC with $3\,\iab$ at $\sqrt{s} = 14 \tev$:\\
solid blue and solid red: di-lepton searches (with soft
leptons)~\cite{ATLAS:2018jjf}. 

\item HE-LHC with $15\,\iab$ at $\sqrt{s} = 27 \tev$:\\
red dashed: same analysis as in \citere{ATLAS:2018jjf}, but rescaled
by~1.5 to take into account the higher energy~\cite{CidVidal:2018eel}. 

\item FCC-hh with $30\,\iab$ at $\sqrt{s} = 100 \tev$:\\
magenta dashed: same analysis as in \citere{ATLAS:2018jjf}, but
rescaled to take into account the higher energy~\cite{Strategy:2019vxc}.\\
solid magenta: mono-jet searches for very soft $\cha1$
decays~\cite{CidVidal:2018eel,Golling:2016gvc} (see also the discussion
in \citere{Berggren:2020tle}), covering
$\De m = 0.7 \gev \ldots 1 \gev$. The upper limit can extend to higher
values. 

\item ILC with $0.5\,\iab$ at $\sqrt{s} = 500 \gev$ (ILC500):\\
solid light green: $\cha1\cha1$ or $\neu2\neu1$ production, which is
sensitive up to the kinematic limit~\cite{Berggren:2020tle} (and
references therein).

\item ILC with $1\,\iab$ at $\sqrt{s} = 1000 \gev$ (ILC1000):\\
gray dashed: $\cha1\cha1$ or $\neu2\neu1$ production, which is
sensitive up to the kinematic limit~\cite{Berggren:2020tle} (and
refrences therein).

\item CLIC with $1\,\iab$ at $\sqrt{s} = 380 \gev$ (CLIC380):\\
very dark green dot-dashed: $\cha1\cha1$ or $\neu2\neu1$ production, which is
sensitive up to the kinematic limit~\cite{Strategy:2019vxc} (see
also \citere{Berggren:2020tle}). 

\item CLIC with $2.5\,\iab$ at $\sqrt{s} = 1500 \gev$ (CLIC1500):\\
green dot-dashed: $\cha1\cha1$ or $\neu2\neu1$ production, which is
sensitive nearly up to the kinematic limit~\cite{Strategy:2019vxc} (see
also \citere{Berggren:2020tle}).

\item CLIC with $5\,\iab$ at $\sqrt{s} = 3000 \gev$ (CLIC3000):\\
dark green dot-dashed: $\cha1\cha1$ or $\neu2\neu1$ production, which is
sensitive nearly up to the kinematic limit~\cite{Strategy:2019vxc} (see
also \citere{Berggren:2020tle}).

\end{itemize}

Not shown in \reffi{fig:future} are searches for disappearing tracks,
which are relevant for very small mass differences,
see \reffis{mc1-delm-wino} and \ref{mc1-tau-wino}. They cut out the
lower edge of the dark blue points, wino DM. The life-time limits are
expected to improve at the HL-LHC by a factor of $\sim
2$~\cite{CidVidal:2018eel}, only slightly cutting into the still allowed
region.

In the higgsino DM case (blue) it can be seen that the HL-LHC can cover
part of the allowed parameter space, but that a full coverage can only
be reached at a high-energy $e^+e^-$ collider with
$\sqrt{s} \lsim 1000 \gev$ (i.e.\ ILC1000 or CLIC1500). The wino DM
case (dark blue), has larger production cross sections at
$pp$~colliders. However, the $\De m$ is so small in this scenario that
it largely escapes the HL-LHC searches (and the different
production cross section turns out to be irrelevant). It can be 
expected that $\De m \gsim 0.7 \gev$ can be covered with the monojet
searches at the FCC-hh. Very low, but still allowed  mass differences
can be covered by the disappearing track searches. However, as for the
higgsino DM case, also here a high-energy $e^+e^-$ collider will be
necessary to cover the whole allowed parameter space. While the
currently allowed points would require CLIC1500, a parameter space
reduced by the HL-LHC disappearing track searches, resulting e.g.\ in
$\mneu1 \lsim 500 \gev$ could be covered by the ILC1000.

The more complicated case for the future collider analysis is given by
the bino/wino parameter points (turquoise), since the limits shown not
only assume a 
small mass difference between $\neu1$ and $\cha1$, but also $pp$ production
cross sections as for the higgsino case. For bino/wino DM these
production cross sections turn out to be larger (as for the pure
wino case), i.e.\ 
displaying the wino/bino points in \reffi{fig:future} should be regarded
as conservative with respect to the $pp$ based limits. Consequently, it
is expected that the HE-LHC or ``latest'' the FCC-hh would cover this
scenario entirely. On the other hand, the $e^+e^-$ limits should be
directly applicable, and large parts of the parameter space will be
covered by the ILC1000, and the entire parameter space by CLIC1500.

\medskip
\noindent
As discussed in \refse{sec:scan} we have not considered explicitly the
possibility of $Z$~or $h$~pole annihilation to find agreement of the
relic DM density with the other experimental measurements. 
However, it should be noted that in this context an LSP with
$M \sim \mneu1 \sim \MZ/2$ or $\sim \Mh/2$ (with $M = M_1$ or $M_2$ or $\mu$)
would yield a detectable
cross-section $e^+e^- \to \neu1\neu1\ga$ in any future high-energy $e^+e^-$
collider. Furthermore, in the case of higgsino or wino DM, 
this scenario automatically yields other clearly detectable EW-SUSY
production cross-sections at future $e^+e^-$ colliders. For bino/wino DM
this would depend on the values of $M_2$ and/or $\mu$. We leave this
possibility for future studies.

On the other hand, the possibility of $A$-pole annihilation was
discussed for all five scenarios. While it appears a rather remote
possibility (particularly for higgsino and wino DM),
it cannot be fully excluded by our analysis. However,
even in the ``worst'' case of $\Slpm$-coannihilation Case-L an upper
limit on $\mneu1$ of $\sim 300 \gev$ can be set (about $\sim 40 \gev$ as
as in the case with the relic DM density taken as a direct
measurement~\cite{CHSold}. 
While not as low as
in the case of $Z$~or $h$-pole annihilation, this would still offer
good prospects for future $e^+e^-$ colliders. We leave also this
possibility for future studies.


\section {Conclusions}
\label{sec:conclusion}

The electroweak (EW) sector of the MSSM, consisting of charginos,
neutralinos and scalar leptons can account for a variety of experimental
data. Concerning the CDM relic abundance, where the MSSM
offers a natural candidate, the lightest neutralino,~$\neu1$,
while satisfying the (negative) bounds from
DD experiments. Concerning the LHC searches, because 
of comparatively small EW production cross-sections, a relatively light
EW sector of the MSSM is also in agreement with
the latest experimental exclusion limits.
Most importantly, the EW sector of the
MSSM can account for the long-standing $3-4\,\sig$ discrepancy of \gmin2.
Improved experimental results are expected soon~\cite{gmt-new} by the 
publication of the Run~1 data of the ``MUON G-2'' experiment.

In this paper we assume that the $\neu1$ provides MSSM DM candidate,
where we take the DM relic abundance measurements~\cite{Planck}
as an upper limit. We analyzed several SUSY scenarios, scanning the EW
sector (chargino, neutralino and slepton masses as well as $\tb$),
taking into account all
relevant experimental data: the current limit for \gmin2\,, the relic
density bound (as an upper limit), the DD experimental bounds, as well
as the LHC searches for EW SUSY particles.
Concerning the latter we included all relevant
existing data, mostly relying on re-casting via \CM, where several
channels were newly implemented into \CM~\cite{CHSold}.

Concretely, 
we analyzed five scenarios, depending on the hierarchy between $M_1$,
$M_2$ and $\mu$ as well as the mechanism that brings the
relic density in agreement with the experimental data.
For $\mu < M_1, M_2$ we have higgsino DM, for $M_2 < M_1, \mu$ wino DM,
whereas for $M_1 < M_2, \mu$ one can have mixed bino/wino DM if 
$\cha1$-coannihilation is responsible for the correct relic density
(i.e.\ $M_1 \lsim M_2$), or one can have bino DM if
$\Slpm$-coannihilation yields the correct relic density, with the mass of the
``left-handed'' (``right-handed'') slepton close to $\mneu1$, Case-L
(Case-R).
Our scans naturally also consist ``mixed'' cases. Higgsino and wino DM,
together with the \gmin2\ constraint can only be fulfilled if the 
relic density is substantially smaller than the measured density. The
other three scenarios can easily accommodate the relic DM density as
direct measurement, which was analyzed in \citere{CHSold}. In the
present analysis these three scenarios were extended to the case where
the relic density is used only as an upper bound. 

We find in all five cases a clear upper limit on $\mneu1$.
These are $\sim 500 \gev$ for higgsino DM,
$\sim 600 \gev$ for wino DM, 
while for $\cha1$-coannihilation we find $\sim 600 \gev$, for
$\Slpm$-coannihilation Case-L $\sim 500 \gev$ and for Case-R
values up to $\sim 500 \gev$ are allowed. Similarly,
upper limits to masses of the coannihilating SUSY particles are found as,
$\mneu2 \sim \mcha1 \lsim 510 \gev$ for higgsino DM,
$\mcha1 \lsim 600 \gev$ for wino DM, 
$\mcha1 \lsim 650 \gev$ for bino/wino DM with $\cha1$-coannihilation and  
$\msl{L} \lsim 600 \gev$ for bino DM case-L and 
$\msl{R} \lsim 600 \gev$ for bino DM case-R.
For the latter, in the
$\Slpm$-coannihilation case-R, the upper limit on the
lighter $\stau$ is even lower, $\mstau2 \lsim 500 \gev$.
The current \gmin2\ constraint also yields limits on the rest of the EW
spectrum, although much loser bounds are found.
These upper bounds set clear
collider targets for the HL-LHC and future $e^+e^-$ colliders.

In a second step we assumed that the new result of the
Run~1 of the ``MUON G-2'' collaboration at Fermilab yields a precision
comparable to the existing experimental result with the same central
value. We analyzed the potential impact of the combination of the Run~1
data with the existing result on the allowed MSSM parameter space.  
We find that the upper limits on the LSP masses are decreased
to about $\sim 480 \gev$ for higgsino DM,
$\sim 500 \gev$ for wino DM, as well as 
$\sim 500 \gev$ for $\chapm1$-coannihilation,
$\sim 450 \gev$ for $\Slpm$-coannihilation Case-L
and $\sim 380 \gev$ in Case-R,
sharpening the collider targets substantially.
Similarly, the upper limits on the NLSP masses go down to about
$\mneu2 \sim \mcha1 \lsim 490 \gev$ for higgsino DM,
$\mcha1 \lsim 600 \gev$ for wino DM, 
$\mcha1 \sim 550 \gev$ for bino/wino DM with $\cha1$-coannihilation,
as well as 
$\msl{L} \lsim 550 \gev$ for bino DM case-L and
$\msl{R} \lsim 440 \gev$, $\mstau2 \lsim 380 \gev$ for bino DM case-R. 

For the three cases with a small mass difference between the lightest
chargino and the LSP (higgsino DM, wino DM and bino/wino DM with
$\cha1$-coannihilation) we have also briefly analyzed the prospects for
future collider searches, specially targeting these small mass
differences. The results for the points surviving all (current)
constraints have been displayed in the plane of the Next-to-LSP (in
these cases $\mcha1$) vs.\ $\De m = \mcha1 - \mneu1$, overlaid with the
anticipated limits from future collider searches from HL-LHC, HE-LHC,
FCC-hh, ILC and CLIC~\cite{Strategy:2019vxc,Berggren:2020tle}.
(These limits are directly applicable to higgsino and wino DM, and
likely to be overly conservative for bino/wino DM with
$\cha1$-coannihilation.) 
While parts of the parameter spaces can be covered by future $pp$
machines, a full exploration of the parameter space requires a future
$e^+e^-$ colliders. Taking into account limits from future $e^+e^-$
machines, a center-of-mass energy of $\sqrt{s} = 1 \tev$ will be
sufficient to conclusively explore higgsino DM, wino DM and bino/wino DM
with $\cha1$-coannihilation.

\medskip
While we have attempted to cover nearly the full set of possibilities
that the EW spectrum of the MSSM presents, while being in agreement with
all the various experimental constraints, our
studies can be extended/completed in the following ways. One can analyze
the cases of:
(i) complex parameters in the chargino/neutralino sector (then also
taking EDM constraints into account);
(ii) different soft SUSY-breaking parameters in the three generations of
sleptons, and/or between the left- and right-handed entries in the case
of $\chapm1$-coannihilation;
(iii) $A$-pole annihilation, in particular in the case of
$\Slpm$-coannihilation for very low $\mneu1$ and $\tb$ values;
(iv) $h$-~and $Z$-pole annihilation, which could be realized for
sufficiently heavy sleptons.
On the other hand, one can also restrict oneself further by assuming
some GUT relations between, in particular, $M_1$ and $M_2$. 
We leave these analyses for future work.

\medskip
In this paper we have analyzed in particular the impact of \gmin2\
measurements on the EW SUSY spectrum. The current measurement sets clear
upper limits on many EW SUSY particle masses. On the other hand we have
also clearly demonstrated the potential of the upcoming measurements of the
``MUON G-2'' collaboration, which have a strong potential of
sharpening the future collider experiment prospects. We are eagerly
awaiting the new ``MUON G-2'' result 
to illuminate further the possibility of relatively light EW BSM
particles.


\subsection*{Acknowledgments}

We thank
M.~Berggren,
J.~List
and
D.~St\"ockinger
for helpful discussions.
We thank C.~Schappacher for the calculation chargino/neutralino mass shifts.
We thank T.~Stefaniak for the evaluation of the latest direct search
limits for heavy MSSM Higgs bosons in the $\Mh^{125}(\tilde\chi)$
scenario~\cite{Bahl:2018zmf}, using {\tt HiggsBounds}~\cite{Bechtle:2008jh,Bechtle:2011sb,Bechtle:2013wla,Bechtle:2015pma,Bechtle:2020pkv}.
I.S.\ gratefully thanks S.~Matsumoto for the cluster facility.
The work of I.S.\ is supported by World Premier
International Research Center Initiative (WPI), MEXT, Japan. 
The work of S.H.\ is supported in part by the
MEINCOP Spain under contract PID2019-110058GB-C21 and in part by
the AEI through the grant IFT Centro de Excelencia Severo Ochoa SEV-2016-0597.
The work of M.C.\ is supported by the project AstroCeNT:
Particle Astrophysics Science and Technology Centre,  carried out within
the International Research Agendas programme of
the Foundation for Polish Science financed by the
European Union under the European Regional Development Fund.



\newcommand\jnl[1]{\textit{\frenchspacing #1}}
\newcommand\vol[1]{\textbf{#1}}

\newpage{\pagestyle{empty}\cleardoublepage}


\end{document}